\documentclass[preprint,11pt]{elsarticle}

\usepackage{graphicx}
\usepackage{import}
\usepackage{blindtext}
\usepackage{titlesec}
\usepackage{enumitem}
\usepackage[table,xcdraw]{xcolor}
\usepackage{tabularx}
\usepackage{booktabs}
\usepackage{ltablex}
\usepackage{todonotes}
\usepackage{hyperref}
\usepackage{float}
\usepackage{booktabs}
\usepackage{lscape}
\usepackage{longtable}
\usepackage{cleveref}
\usepackage{listings}
\usepackage{multirow}
\usepackage{pgfplots}
\pgfplotsset{width=\textwidth,compat=1.8}
\makeatletter
\def\ps@pprintTitle{%
 \let\@oddhead\@empty
 \let\@evenhead\@empty
 \def\@oddfoot{\centerline{\thepage}}%
 \let\@evenfoot\@oddfoot}
\makeatother

\newcommand{\quickwordcount}{%
  \immediate\write18{texcount -1 -sum -merge \jobname.tex > \jobname-words.sum }%
  \input{\jobname-words.sum} words%
}

\definecolor{darkpastelpurple}{rgb}{0.59, 0.44, 0.84}
\definecolor{darkmagenta}{rgb}{0.55, 0.0, 0.55}

\tikzstyle{mybox} = [draw=black, very thick, rectangle, rounded corners, inner ysep=5pt, inner xsep=5pt]
\usepackage{color}
\newcommand{\rf[1]}{\textcolor{black}{#1}}
\newcommand{\rs[1]}{\textcolor{black}{#1}}
\newcommand{\rt[1]}{\textcolor{black}{#1}}
\newcommand{\ra[1]}{\textcolor{black}{#1}}

\begin{document}
\begin{frontmatter}
\title{Motivations, Benefits, and Issues for Adopting Micro-Frontends: A Multivocal Literature Review}

\author[TUNI]{Severi Peltonen}
\ead{severi.peltonen@gmail.com }

\author[DAZN]{Luca Mezzalira}
\ead{luca.mezzalira@dazn.com}

\author[TUNI]{Davide Taibi}
\ead{davide.taibi@tuni.fi}

\address[TUNI]{Tampere University, Tampere (Finland)}
\address[DAZN]{DAZN. London, United Kingdom\footnote{\textbf{Please cite as: }
Severi Peltonen, Luca Mezzalira, Davide Taibi. Motivations, Benefits, and Issues for Adopting Micro-Frontends: A Multivocal Literature Review. 
Information and Software Technology - In Press (2020)}}

\begin{abstract}

[Context]
Micro-Frontends are increasing in popularity, being adopted by several large companies, such as DAZN, Ikea, Starbucks and may others. Micro-Frontends enable splitting of monolithic frontends into independent and smaller micro applications. However, many companies are still hesitant to adopt Micro-Frontends, due to the lack of knowledge concerning their benefits. Additionally, provided online documentation is often times perplexed and contradictory.

[Objective] The goal of this work is to map the existing knowledge on Micro-Frontends, by understanding the motivations of companies when adopting such applications as well as possible benefits and issues.

[Method] For this purpose, we surveyed the academic and grey literature by means of the Multivocal Literature Review process, analyzing 173 sources, of which 43 reported motivations, benefits and issues. 

[Results] The results show that existing architectural options to build web applications are cumbersome if the application and development team grows, and if multiple teams need to develop the same frontend application.  In such cases, companies adopted Micro-Frontends to increase team independence and to reduce the overall complexity of the frontend. The application of the Micro-Frontend, confirmed the expected benefits, and Micro-Frontends resulted to provide the same benefits as microservices on the back end side, combining the development team into a fully cross-functional development team that can scale processes when needed.
However, Micro-Frontends also showed some issues, such as the increased payload size of the application, increased code duplication and coupling between teams, and monitoring complexity. 

[Conclusions] Micro-Frontends allow companies to scale development according to business needs in the same way microservices do with the back end side. In addition, Micro-Frontends have a lot of overhead and require careful planning if an advantage is achieved by using Micro-Frontends. Further research is needed to carefully investigate this new hype, by helping practitioners to understand how to use Micro-Frontends as well as understand in which contexts they are the most beneficial.  

\end{abstract}


\end{frontmatter}

\section{Introduction}

Developing the presentation layer of a modern web application has become a major and crucial task for industrial companies. Development teams are constantly looking for new ways to develop, deploy, and maintain applications in an effective manner so companies can quickly and effectively deliver value for their customers. 

New front-end frameworks are continuously introduced into the market and developers have many valid options to build powerful feature-rich web applications such as single-page application (SPA), server-side rendering application (SSR), or static HTML files combined to a web page. However,  most of them end up being monolith front-ends.
Hence, the client-side of the application grows, and its development becomes hard to scale, especially if different teams need to edit the same front-end application simultaneously. 

Micro-Frontends~\cite{thoughtworksTechRadar}\cite{MezzaliraBook}\cite{Soderlund}\cite{GeersMichael}
were introduced in 2016~\cite{thoughtworksTechRadar} to enable the decomposition of the front-end into individual and semi-independent front-ends, separating the business logic from the frontend, and creating independent services that interact together~\cite{MezzaliraIEEESW2020}.  
Micro-Frontends are nowadays adopted by several large industries including DAZN, Ikea, New Relic, SAP, Springer, Starbucks, Zalando,    and many others. 

Micro-Frontends share the main principles, benefits, and issues of microservices~\cite{thoughtworksTechRadar}: both are modelled around business domains, hiding implementation details between them. Each team should own its microservice (back-end) and the related frontend, enabling to decentralize decisions and deploy independently.     
However, Micro-Frontends also introduce some drawbacks, such as the risk of communication overhead if the system is not well designed and revised with the business growth, potential performance issues when the vendors of a project are not carefully taken into account (for instance,  we do for SPA), and broken user experience when the governance behind a design system is not well thought out.

Different software architects are pushing for this architectural style at practitioner forums. However, considering the costs some practitioners are still hesitant to adopt Micro-Frontends, because they are not fully aware of the pros and cons.

So, software developers often choose to adopt one architecture over another based on their experience in previous projects or based on the perceived benefits of the new architecture. Therefore, it is important to study why Micro-Frontends have been adopted, to understand the current motivations behind their adoption, and to investigate whether specific issues are believed to require more improvement than others.
To elicit these motivations, we conducted an empirical study in the form of a Multivocal Literature Review (MLR)~\cite{Garousi18a}.

Therefore, the contribution of this work, is aimed to identify: 
\begin{itemize}
    \item the motivations that led practitioners to adopt Micro-Frontends
    \item the benefits achieved by the companies that adopted Micro-Frontends
    \item the issues that practitioners experienced 
\end{itemize}

To the best of our knowledge, only a limited number of studies have investigated Micro-Frontends~\cite{Yang_2018}\cite{Menaetal19}.
This work will help companies to understand how Micro-Frontends can be beneficial for their needs, and if motivations, issues and benefits that other companies experienced match their expectancy. Moreover, this work can help researchers to understand the new trend, while at the same time opening up new avenues for future research on web front-ends.



The remainder of this paper is structured as follows. Section~\ref{sec:Background} presents the background of this work, introducing Micro-Frontends, and comparing them with Microservices. Section~\ref{sec:RelatedWork} discusses the related works on Micro-Frontends. Section~\ref{sec:ResearchQuestions} describes the Research Questions we proposed while Section~\ref{sec:design} provides detailed information on the MLR process we adopted. Section~\ref{sec:results} reports the results to our RQs and discusses them. Section~\ref{sec:discussion} discusses implications for practitioners and researchers. Section~\ref{sec:ttv} highlights the threats to validity while finally, Section~\ref{sec:conclusion} draws the conclusions. 
\label{sec:Introduction}

\section{Background}


\rs[As development teams start to design and create new web applications, there are many different
architectures, approaches and tools that the development team can choose from. 
]

\rs[In this Section, we provide an overview of the alternative options for developing an frontend applications and how these options differ from Micro-Frontends.
This section also provides an theoretical background for the Micro-Frontends and how Micro-Frontends are related to Microservice architecture.]

\subsection{JAMstack Architecture}

\rs[JAMstack stands for JavaScript, APIs and Markup, and  is an increasingly popular web development philosophy that aims to speed up both the web development process and webpage download times \cite{FruhlingerJamstack}.]

\rs[JAMstack is an architecture for building modern web pages and applications based on advanced tools and workflows for a faster, simpler and more secure web. It delivers the speed and simplicity of pre-rendered static sites with dynamic capabilities via JavaScript, APIs and serverless functions. These pre-rendered sites are deployed directly to CDN without requiring to manage, scale, or patch any web servers. For JAMstack there is no server]

\begin{figure}[ht!]
  \includegraphics[width=\textwidth]{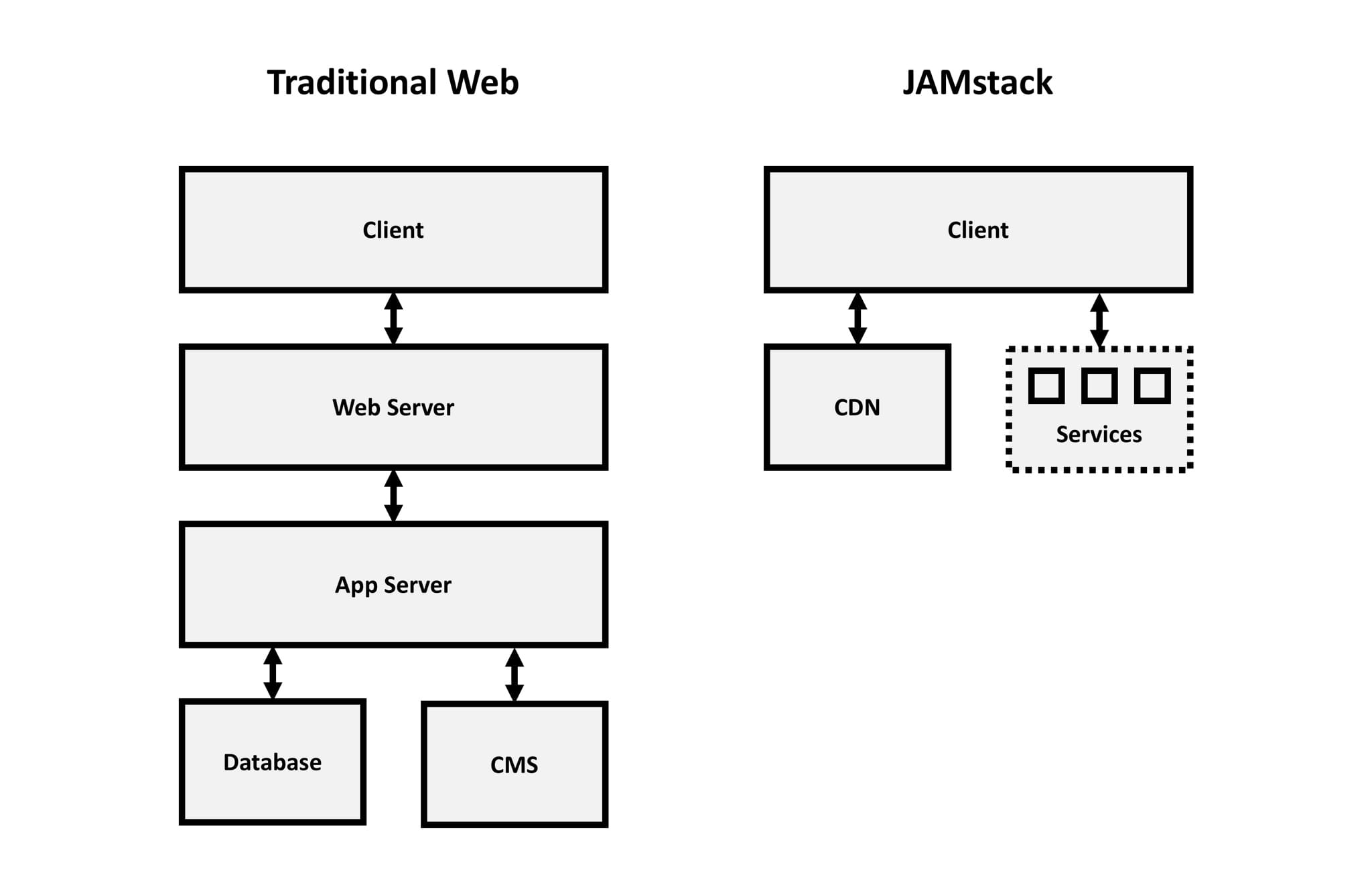}
  \caption{The differences between traditional web architecture and JAMstack web architecture.}
  \label{fig:jamstack}
\end{figure}

\noindent \rs[environment at all. Therefore,  JAMstack applications are less expensive, because hosting static files is cheap. This also means that frontend developers can focus only on the frontend development and debugging, this usually means a more focused approach on the final result \cite{MezzaliraBook}.]
 
 \rs[Figure \ref{fig:jamstack} shows the main differences between traditional monolithic web application and JAMstack architecture based application. Throughout this work, there is a lot of discussion on breaking the monolith, which means to break the frontend application into smaller parts, which is the case with Micro-Frontends. In case of the JAMstack architecture on the other hand, it centralizes more on separating the frontend and back-end completely from each other.]

\subsection{Client-Side Rendering application}
\label{sec:ClientSideRendering}
\rs[A frontend application can also be a client-based where a client (browser) receives HTML and JavaScript files, that manipulate the HTML document and the DOM. This results in interactive applications where user interactions result in animations or GUI changes, without a need of page refresh. To facilitate this, browser exposes an API, called Document Object Model (DOM), that allows scripting languages to access and manipulate HTML documents~\cite{DomCSR}. This manipulation is most commonly done using JavaScript.]

  \rs[CSR can even be used to render all of the content of a web page, and even simulate page navigation by re-rendering most or all of the web page. The result can be a native-like experience to the user \cite{MixuSPA}. This type of frontend application is called a Single-Page Application, or SPA in short. SPA is a web application which is delivered to the browser in a single HTML file and it uses CSR to change the content which will be shown to user \cite{SPA}.]

\subsubsection{Single-page Application}

\begin{figure}[H]
  \includegraphics[width=\textwidth]{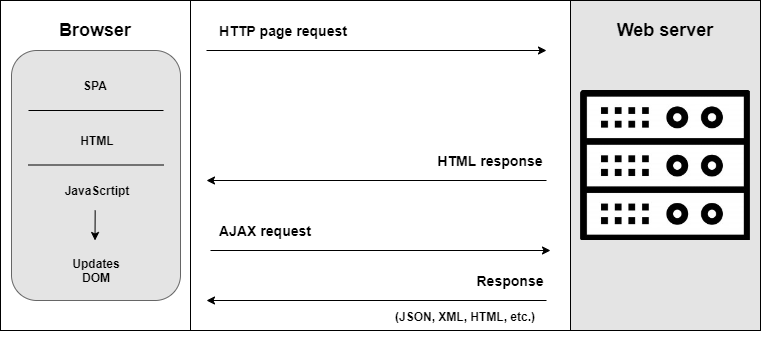}
  \caption{Client-side rendering concept.}
  \label{fig:ClientSide}
\end{figure}

\rs[In Single-page Applications (SPA), only one HTML file is loaded to the browser, the page does
not need to refresh when the user interacts with the page. This gives the user a smoother
user experience.
To be able to change the content of the page and to provide additional information for the
user, the DOM  has to be dynamically updated by using JavaScript
and HTTP requests to get information from the server \cite{w3schools}. There is a large number of
frameworks, libraries and tools, for example, Angular, React and Vue, which are designed to
create a layer of abstraction for the DOM manipulation process \cite{AngularSPA}\cite{ReactSpa}\cite{VueJs}. However,
the given example SPA frameworks are not compatible with each other as such, the core idea behind
all of them are the same, see Figure \ref{fig:ClientSide}.]

\rs[As the single HTML file is loaded by the browser, the HTML provides a rooting point for the JavaScript
application, which is loaded besides with the HTML file. Also images,
CSS, script files, and other external resources are loaded \cite{jadhavAngular}.
The rooting point is basically a single HTML element, most typically a block \textless div\textgreater -element. 
When the rooting point is provided for the JavaScript application, the applications knows where to start compiling HTML
content to the document \cite{jadhavAngular}.]

\rs[All frameworks developed for creating single-page applications have application life-cycle
methods with standardized names which give the developers the ability to define what will
be changed in the DOM and when during application life cycle ~\cite{ReactJsLifeCycle} \cite{VueJsLifeCycle} \cite{AngularLifeCycle}. 
This is one of the benefits of the SPA. The code is downloaded only once at the beginning of the application life-cycle and the entire logic is available upfront \cite{MezzaliraBook}.]

  \rs[In short, there is a method which defines
what happens when the application component will be displayed for the user and
another method determining what happens when the application component will
be removed from the view. As the SPA avoids multiple network calls for loading additional application logic and renders right content instantaneously during the application life cycle, user experience is enhanced and application is able to simulate native-like applications \cite{MezzaliraBook}.]

\rs[Even though SPA is nowadays a popular way to create web applications it has some disadvantages for certain type of applications.
The initial application is loaded when the application starts and this takes usually longer time than with other architectures because the whole application needs to be loaded to the browser instead of only what the user needs to see at that time \cite{MezzaliraBook}.]

\subsection{Isomoprhic Applications}

\rs[All different frontend applications methods can be mixed, and different parts of the HTML document can be rendered using client-side or server-side methods. The main idea of isomorphic web applications, or universal applications, is to write JavaScript applications designed for the web browser but at the same time, the application must run on the server for generating HTML markup files. So, the code between the server and the client is shared and can run in both context \cite{MezzaliraBook}.]

\rs[This technique brings some benefits when used in the right way. It is in particular convenient when the time to interaction, A/B testing, and SEO are essential characteristics for the application \cite{MezzaliraBook}.]

\rs[The isomorphic application can be designed in different ways but
the main concept is that a web page is rendered twice by the same application. Because the web application share code between client and server, the server, for instance can do the rendering part for the page requested by the browser, retrieve the data to display from the database or from one or multiple APIs, compile the data together, and then pre-render it with the template system used for generating the view, in order to serve to the client a page that doesn’t need additional network calls for requesting additional data to display. The benefit compared to a traditional SPA is that the web page is loaded quicker, as the initial file contains all required HTML to show the web page \cite{MediumIsomorphic}. Although, with Isomorphic application time-to-interaction is larger, because there will be an amount of time where user sees graphical elements on the screen that appears interactive but are not. This is called \textit{Uncanny valley}, where browser has rendered the servers response and then JavaScript is downloaded, parsed and executed \cite{KotteIsomorphic}.]

\rs[As universal application uses both frontend and back-end for generating views for user, these applications could suffer from scalability problems if the web page is visited by millions of users as the application might pre-render HTML pages on the server \cite{MezzaliraBook}. Micro-Frontends can also suffer from this problem if server-side composition is used because a lot of different micro applications needs to be stitched together on server.]

\subsection{Static HTML sites}

\rs[In the end, everything comes out as HTML on the frontend.
Web pages are written in HTML, the declarative web programming language that tells web browsers how to structure and present content on a web page. In other words, HTML provides the basic building blocks for the web. And for a long time, those building blocks were pretty simple and static: lines of text, links and images. An HTML site is still quite common, and may be the right solution for a small site or start-up business landing page. Static HTML site are not meant to be a long lasting or at least developer cannot create full application with it.]

\subsection{Micro-Frontends}
\label{subsec:MicroFrontendOverview}
No silver bullet for designing a software architecture has been made and there will be no such thing coming in the future. Nonetheless, software development practitioners and researchers are constantly searching and developing new ways to create software applications which are fast to develop and deploy as well as easy to maintain. Also, companies have to adopt new agile methodologies into their organizational structures to respond to customer needs at unprecedented speeds \cite{Bosch16a}.

Working on the frontend side of the application developers and software architects have a few architectural options to choose from e.g., single-page applications, SPAs, in short, server-side rendering applications, or application composed by static HTML files. Over time these architectures might lead the project to become monoliths. This increases the complexity of the frontend application and making changes on part of the system may have unnecessary or unwanted effects on other parts. Code bases become huge, the application has a lot of dependencies and becomes tightly coupled, coordination between development teams becomes harder and slower, which leads to the law of diminishing returns. Increasing the number of developers on frontend teams will not affect the production rate, since the chosen architecture has set boundaries for developers.

Micro-Frontends extends the concepts of Microservices to the frontend side of the application. It transforms monolithic web applications from a single code based application architecture to an application that
combines multiple small frontend applications into one whole. Each of these independent applications can run, and be developed and deployed independently. The capability of independent development and deployment allows development teams to build isolated and loosely coupled services.
The idea behind Micro-Frontends is to handle a web application as a combination of features or business sub-domains. Each team should have only one domain to handle. 

\begin{figure}[H]
  \includegraphics[width=\textwidth]{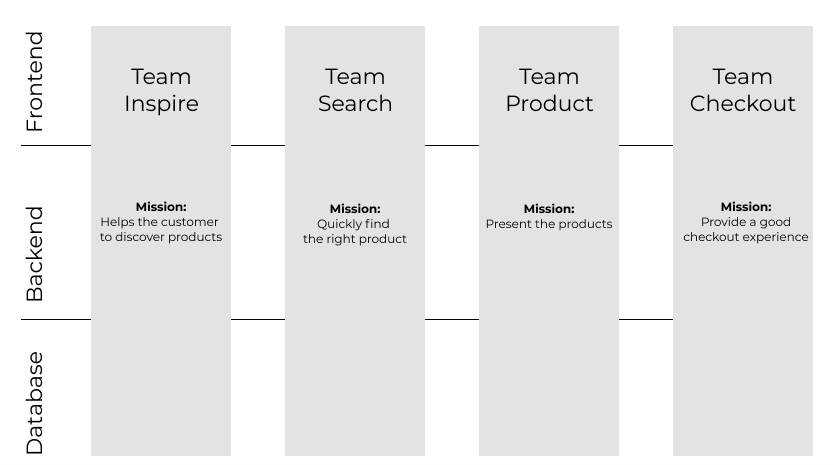}
  \caption{End-to-end frontend teams with Micro-Frontends architecture.}
  \label{fig:VerticalTeams}
\end{figure}

Frontend monoliths introduce horizontal layers to the frontend side of the application, but Micro-Frontends aim to divide the application vertically as shown in Figure~\ref{fig:VerticalTeams}. 
Each of these vertical slices serves a specific business domain or feature and is built completely from the bottom to the top. With Micro-Frontends, each development team can be technologically agnostic and decide what kind of technology stack to use. Teams can update or even switch the stack without cooperating with other teams.

\subsection{Micro-Frontends Composition}

For architecting a Micro-Frontends application there are a few different options to choose from. With Micro-Frontends architecture some architectural decisions have to be made upfront because these decisions will shape the future decisions which are done alongside the project.

To define Micro-Frontends, the key decision to make is a need to identify how to consider a Micro-Frontend from the technical point-of-view. For this there are two options:

\begin{itemize}
\item \textit{Horizontal split: multiple Micro-Frontends per page}
\item \textit{Vertical split: one Micro-Frontend per time}
\end{itemize}

In a horizontal split, multiple smaller applications are loaded to the same page and this requires that multiple teams need to coordinate their efforts since each team is responsible for a part of the view. 

With the vertical split scenario, each team is responsible for a business domain e.g. authentication or payment experience. With the vertical split, Domain-Driven Design (DDD) will apply.

\rt[
In case of horizontal split, an important step is to define how micro-frontends communicate with each other. One method is to use an event emitter injected into each micro-frontend. This would make each micro-frontend totally unaware of its fellows, and make it possible to deploy them independently. When a micro-frontend emits an event, the other micro-frontends subscribed to that specific event react appropriately.
It is also possible to use custom events. These have to bubble up to the window level in order to be heard by other micro-frontends, which means that all micro-frontends are listening to all events happening within the window object. They also dispatch events directly to the window object, or bubble the event to the window object, in order to communicate.]

\rt[In case of a vertical split, it is important to   understand how to share information across micro-frontends. For both horizontal and vertical approaches, we need to think about how views communicate when they change. It’s possible that variables may be passed via query string, or by using the URL to pass a small amount of data (and forcing the new view to retrieve some information from the server). Alternatively, it is possible to use web storage to temporarily (session storage) or permanently (local storage) store the information to be shared with other micro-frontends.
]

\begin{figure}[H]
  \includegraphics[width=\textwidth]{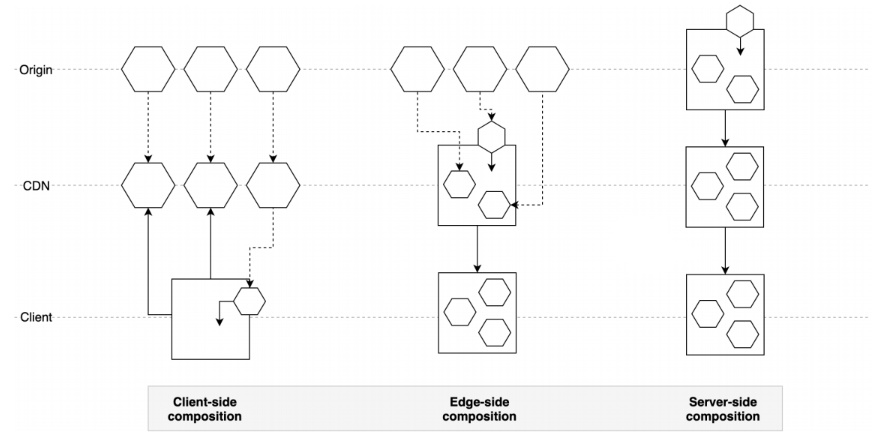}
  \caption{Different ways to combine a Micro-Frontends architecture.}
  \label{fig:mfComposition}
\end{figure}

Three different approaches can be used for composing Micro-Frontend applications(Figure~\ref{fig:mfComposition}):
\begin{itemize}
\item \textit{Client-side composition}
\item \textit{Edge-side composition}
\item \textit{Server-side composition}
\end{itemize}

\subsubsection{Client-Side Composition}

On the client-side composition, an application shell loads Micro-Frontends inside itself. 
Micro-Frontends should have as an entry point a JavaScript or HTML file so that the application shell can dynamically append the DOM nodes in the case of an HTML file or initialize the JavaScript application when the entry point is a JavaScript file.

Another possible approach is to use a combination of iframes for loading different Micro-Frontends otherwise transclusion mechanism, which could be used on the client-side via a technique called client-side include. This is where the application shell is lazy loading components
inside a container using a placeholder tag, and parses all the placeholders by replacing them with the corresponding component.
This approach brings many options to the table. However, using client-side includes has a different effect than using iFrames.
\\

\rt[\textit{Micro-Frontends with client-side rendering}]

\rt[With client-side rendering the web page can be split into Micro-Frontends with approach as showed in Listing 1 and Listing 2.
In the example in line 11, page.js works as applications shell which combines all the fragments loaded under it and puts the content to the main tag. The application shell and the other applications are run in the browser.]

\begin{lstlisting}[language=html, emph={2} 
  , numberstyle=\tiny, stepnumber=2, numbersep=5pt
  , caption={Example of client-side rendering Micro-Frontends.}
  , captionpos=b, nolol=false, label={code:sort}]
<!doctype html>
<html lang="en">
  <head>
    <meta charset="utf-8">
    <title>Client-side rendering Micro-Frontend</title>
    <link href="./page.css" rel="stylesheet">
  </head>
  <body>
    <main id="app"></main>
    <script src="./team-appshell/page.js" async></script>
    <script src="./team-catalog/fragments.js" async></script>
    <script src="./team-actions/fragments.js" async></script>
  </body>
</html>
\end{lstlisting}

\begin{lstlisting}[language=html, emph={2} 
  , numberstyle=\tiny, stepnumber=2, numbersep=5pt
  , caption={Client-side rendering code to render Micro-Frontends.}
  , captionpos=b, nolol=false, label={code:sort1}]
const $app = document.getElementById('app');

function renderPage() {
  $app.innerHTML = `
    <h1 id="heading">The Micro-Frontend</h1>
    <team-catalog id="catalog"></team-catalog>
    <team-actions id="actions"></team-actions>
  `;
}
\end{lstlisting}

\subsubsection{Edge-Side Composition}

With edge-side composition, the web page is assembled at the CDN level. Many CDN providers give us the option of using an XML-based markup language called Edge Side
Include (ESI). ESI is not a new language; it was proposed as a standard by Akamai and Oracle \cite{AkamaiOracle}, among others, in 2001. The reason behind ESI was the possibility of scaling a web infrastructure to exploit the large number of points of presence around the
world provided by a CDN network, compared to the limited amount of data centre capacity on which most software is normally hosted. One of the drawbacks of this implementation is that ESI is not implemented in the same way by each CDN provider; therefore, a multi-CDN strategy, as well as porting application code from one provider
to another, could result in a lot of refactors and potentially new logic to implement.

\subsubsection{Server-Side Composition}

On the server-side composition, which could happen at
runtime or at compile time. In this case, the origin server is composing the view by
retrieving all the different Micro-Frontends and assembling the final page. If the page is highly cacheable, it will then be served by the CDN with a long time-to-live policy;
instead, if the page is personalized per user, it will require serious consideration regarding the scalability of the eventual solution, when there are many requests coming from different clients. When the server-side composition is decided to use, use cases in the application need to be analysed deeply. If runtime composition is used, the project must have a clear scalability strategy for servers in
order to avoid downtime for our users.
After understanding all the possibilities, the decision needs to be made, which technique is more suitable for the project and the structure of the development team. Also, a mix of approaches can be used.

\hfill \break
\noindent\rt[\textit{Micro-Frontends with server-side rendering}]

\begin{lstlisting}[language=html, emph={1} 
  , numberstyle=\tiny, stepnumber=2, numbersep=5pt
  , caption={Example of Server-side rendering Micro-Frontends.}
  , captionpos=b, nolol=false, label={code:Server-side-rendering}
]
export default function renderView() {
return `
<h1 id="heading">The Micro-Frontend</h1>
<team-catalog id="catalog">
<!--#include virtual="/team-catalog" -->
</team-catalog>
<team-actions id="actions">
    <!--#include virtual="/team-actions" -->
</team-actions>
 `;
}
\end{lstlisting}

\begin{figure}[H]

\begin{lstlisting}[language=html, emph={1} 
  , numberstyle=\tiny, stepnumber=2, numbersep=5pt
  , caption={Example of proxy server config.}
  , captionpos=b, nolol=false, label={code:proxy_server_config}
]
upstream catalog {
  server team_catalog:8081;
}
upstream actions {
  server team_actions:8082;
}

server {
  listen 8080;
  ssi on;

  location /catalog {
    proxy_pass  http://team_catalog;
  }
  location /actions {
    proxy_pass  http://team_actions;
  }
  location /shell {
    proxy_pass  http://team_shell;
  }
  location / {
    proxy_pass  http://team_shell;
  }
}
\end{lstlisting}

\end{figure}

\rt[With server-side rendering The web server replaces the directive, for example ``  \textless!--\#include virtual="/team-catalog" --\textgreater`` with the contents of the referenced URL before it passes the markup to the client. The referenced URLs can be defined in the server configuration as shown in Listing 4. Shown example uses server-side include (SSI) method but there are also other possibilities for example Edge-side include (ESI) method which is also valid option for server-side rendering Micro-Frontends.]

\subsection{Microservices vs. Micro-Frontends}

In recent years Microservices have received great attention in the academic field, and have become one of the key research objects in the field of information science, but also in industrial fields where more and more companies are changing their monolithic singe-application back-end implementations to Microservice architectures. As mentioned in section \ref{subsec:MicroFrontendOverview} Micro-Frontends is a fairly new topic in the field of information science, but it has gained significant popularity amongst practitioners in this field.

\begin{figure}[H]
\centering
  \includegraphics[width=0.7\textwidth]{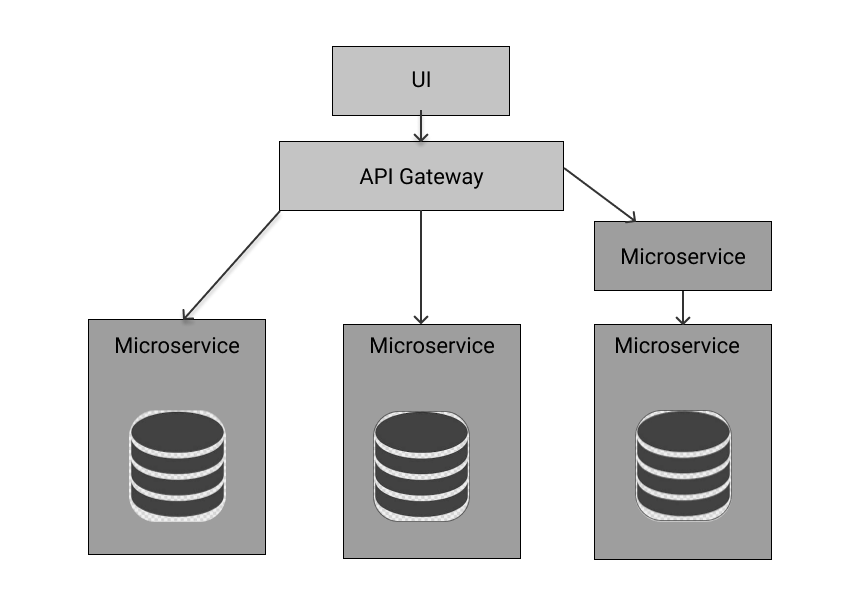}
  \caption{Microservice architecture.}
  \label{fig:Microservices}
\end{figure}

By tradition single application, the system based on a single architectural style contains a large number of modules and dependencies between components, and overtime boundaries between modules become unclear. With tightly coupled modules, modifying one part of the application forces the project to be redeployed entirely. This full redeployment
process takes a long time and has a large impact range. As a tightly coupled application, a single application cannot be targeted for the characteristics of different business modules, and can only be expanded as a whole, resulting in a waste of resources.
\begin{figure}[!hb]
\centering
  \includegraphics[width=0.8\textwidth]{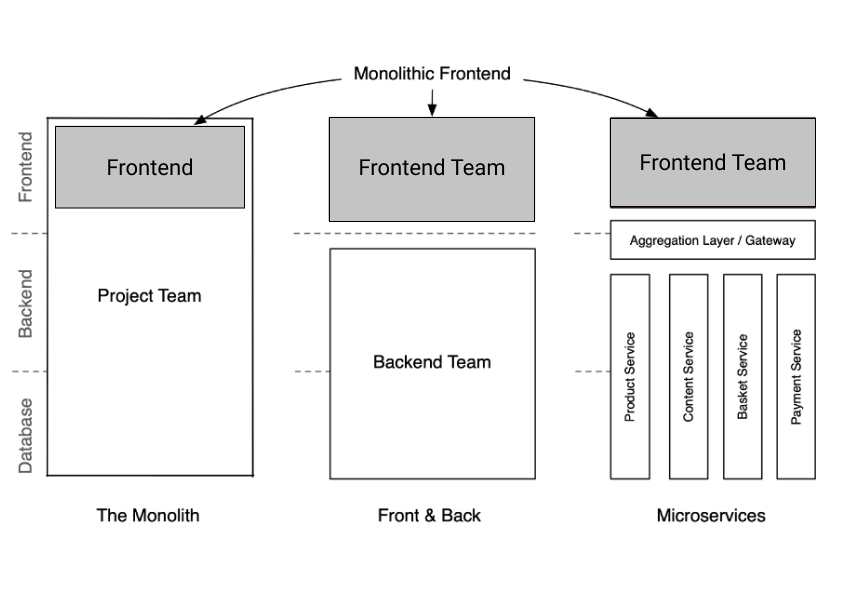}
  \caption{Backend Microservice architecture with a monolithic frontend presentation layer.}
  \label{fig:FrontendMonolith}
\end{figure}
Microservices are a variant of the service-oriented architecture architectural style that builds
applications as a collection of loosely coupled services. It merges complex broad applications in a
modular way based on small functional components that communicate through a collection of language-independent APIs, as shown in figure \ref{fig:Microservices}. Each functional block or service focuses on a single responsibility and function, and can be
developed, tested, and deployed independently \cite{Chen18a}.
This enables development teams to develop applications in parallel. It also enables continuous delivery and deployment.

On the back-end side, the Microservice architecture has relatively mature implementation solutions and its benefits are definite, but the frontend side of the application remains monolithic under current development trend. As mentioned, Micro-Frontends extends the Microservice architecture idea and many principles from Microservices apply to Micro-Frontends:
\begin{itemize}
    \item \textbf{Modeled Around Business Domains}: 
Following the Domain-Driven Design principles (DDD), each piece of the software should align business and technical concerns through analysis of the business domain, modelling of the problem domain \cite{Khononov19a}, and leveraging ubiquitous languages shared across the business.

\item \textbf{Culture of Automation}: 
A strong automation culture allows us to move faster and in a more reliable way. Considering that every Microservice and Micro-Frontends project contains tens if not hundreds of different parts, we need to make sure our continuous integration and continuous deployment pipelines are solid and with a fast feedback loop for embracing this architecture. Investing time to get our automation right will result in the smooth adoption of Microservices but also of Micro-Frontends.
\item \textbf{Hide Implementation Details}: 
Hiding implementation details and programming with contracts are two essential assumptions, especially when parts of the application need to communicate with each other. It’s very important to define a contract upfront between teams and for all parties to respect that during the entire development process. In this way, each team will be able to change the implementation details without impacting other teams, unless there is an API contract change. These practices allow a team to focus on the internal implementation details without disrupting the work of other teams. Because each team can then work at its own pace and without external dependencies, this will result in a more effective integration.
\item \textbf{Decision Decentralization}
Decentralizing the governance empowers developers to take the right decision at the right stage to solve a problem. Often with a monolith architecture, many key decisions are made by the most experienced people in the organization. These decisions, however, often lead to trade-offs alongside the software life-cycle. Decentralizing these decisions could have a positive impact on the entire system by allowing a team to take a technical direction based on the problem(s) they are facing, instead of creating compromises for the entire system. However, it is important that the tech leadership  (architects, principal engineers, CTOs) should provide high-level directions where the team can operate without needing to wait for central decisions.
\item \textbf{Independent Deployment}: One of the benefits of Microservices and Micro-Frontends is the possibility to deploy artefacts independently. Teams can deploy at their own speeds without waiting for external dependencies to be resolved before deploying in production. Considering this with Micro-Frontends and Microservices, it is obvious that a team could own a vertical business domain end to end deciding the best infrastructure, the best frontend and back-end technology suitable for a business domain.
\item \textbf{Failure Isolation}: Considering that, splitting a monolith application into tens, if not, hundreds of services, if one or more Microservices becomes unreachable due to network issues or service failures, the rest of the system ought to be available for users. There are several patterns for providing graceful failures with Microservices and the fact that they are autonomous and independent are just reinforcing the concept of isolated failure. Micro-Frontends require a part of the application to be lazy-loaded or compose a specific view at run-time with the risk to end up with errors due to network failures or 404 not found error. Therefore, the application needs to find a way to avoid impacting the user experience by providing alternative content or just hiding a specific part of the application.
\end{itemize}

\label{sec:Background}

\section{Related Work}
As mentioned earlier, Micro-Frontends 
has not been extensively investigated in research works, mainly because of their novelty.

\rf[Yang et al. \cite{Yang_2018} described a content management system (CMS) created with Micro-Frontends architecture using Mooa Framework \cite{MooaFramework18}.
They reported that frontend and backend application separation provided a
great user experience,
but result in a single page application not being well scaled and deployed. 
They concluded that Micro-Frontend-based CMS design enables teams to develop
independently, quickly deploy and test individually, helping with continuous integration, continuous
deployment, and continuous delivery. Moreover, they also confirmed that Micro-Frontends architecture is still in the adopt stage and is not mature enough.]

\rf[Compared to our work, Yang et al. work was more practical, proposing a small CMS application. Differently from this work, they did not aim at providing overall picture of Micro-Frontends as many of the benefits of the Micro-Frontends. They also created their application with mooa framework which only supports angular 2+ based applications, reducing technology agnosticism of Micro-Frontends. Therefore they do not cover all the possibilities of how Micro-Frontends can be composed. ]



\rf[Mena et al. \cite{Menaetal19} proposed another application of Micro-Frontends, creating a progressive web application by using Microservices and Micro-Frontends architecture. Their  main focus was not applied specifically to Micro-Frontends but more to develop a more generic web application using microservices. However, they also concluded  that Micro-Frontend made possible to build the user interface dynamically and develop visual components independently, finding different ways to show the user data.
Similary to Yang et al. Mena et al.'s approach was practical and this way they can only cover the development team benefits and issues of Micro-Frontends.]


\rf[Pavlenko et al. \cite{Pavlenko2020} proposed another application of Micro-Frontends in the context of Single-page applications. They designed and developed
a SPA frontend application with Micro-Frontends principles, reporting in details on the application design and on the technologies used.]
\rf[Thought, they discover that most of the time their small development team had to focus more on the architecture and development tools rather than focusing on the feature development which increased the overall development time.]
\rf[Moreover, they reported  that Micro-Frontends is not a valid option with smaller development teams or when the seniority of the team is low.  ]

\label{sec:RelatedWork}

\section{Research Questions}

The goal of this work is to systematically map, review, and synthesize state-of-art and -practices in the area of web front-end architectures, so to understand the reasons why this architectural style is getting attention amongst practitioners and industrial companies, also highlighting Micro-Frontends benefits and issues. Moreover, this work also tries to identify opportunities for future research, especially from the point of view of practitioners and industrial companies. 


The novelty of the Micro-Frontend architecture allows us approaching this subject from many different research directions since not much scientific research has been made yet. For this reason, research questions were defined to cover the most basic topics to get a comprehensive view of this subject but not to go too deep into details. 

Based on the aforementioned goal, we formulated three Research Questions (RQs):
\begin{itemize}
  \item [RQ1] Why practitioners are adopting Micro-Frontends?\\
  In this RQ, we aim at understanding the motivations that lead companies to adopt Micro-Frontends for developing web applications. 
  \item [RQ2] What benefits are achieved by using Micro-Frontends? \\
Different software architectures aim to solve different problems that other architectures fail to do or enhance parts of the development process which will eventually affect the life-cycle of the application.
  In this RQ, we want to understand the benefits provided by Micro-Frontends, and which type of problems Micro-Frontends are aimed to solve.

  

\item [RQ3] Do Micro-Frontends introduce any issues?\\
Every technology has benefits and issues. In this RQ we want to understand what issues might occur when using Micro-Frontends and what trade-offs are being made to overcome these issues.
\end{itemize}

\label{sec:ResearchQuestions}

\section{Study Design}
\label{sec:design}

In this Section, we provide an overview of the study process adopted in this work. 
Because of the novelty of the topic, and of the large presence of user-generated content on the web, we adopted a Multivocal Literature Review (MLR) process~\cite{Garousi18a}. 
In the remainder of this Section, we provide an overview of the overall process, the strategy adopted for the search process, the selection, data extraction, and synthesis processes.


\subsection{The MLR Process}
Systematic Multivocal Literature Review (MLR) proved to be the best choice for the research method due to the lack of maturity of the subject. In a normal case, the MLR process is divided so, that it includes both academic and grey literature and the differences between practitioners and academic researchers can be synthesized from the results. 
The key motivation for the inclusion of grey literature is the strong interest of practitioners on the subject and grey literature content creates a foundation for future research. 

We classified peer-reviewed papers as \textit{academic literature}, and other content (blog post, white-papers, Podcasts, ...) as \textit{grey literature}. 

\begin{figure}
  \includegraphics[width=\textwidth, trim={0 300 0 0},clip]{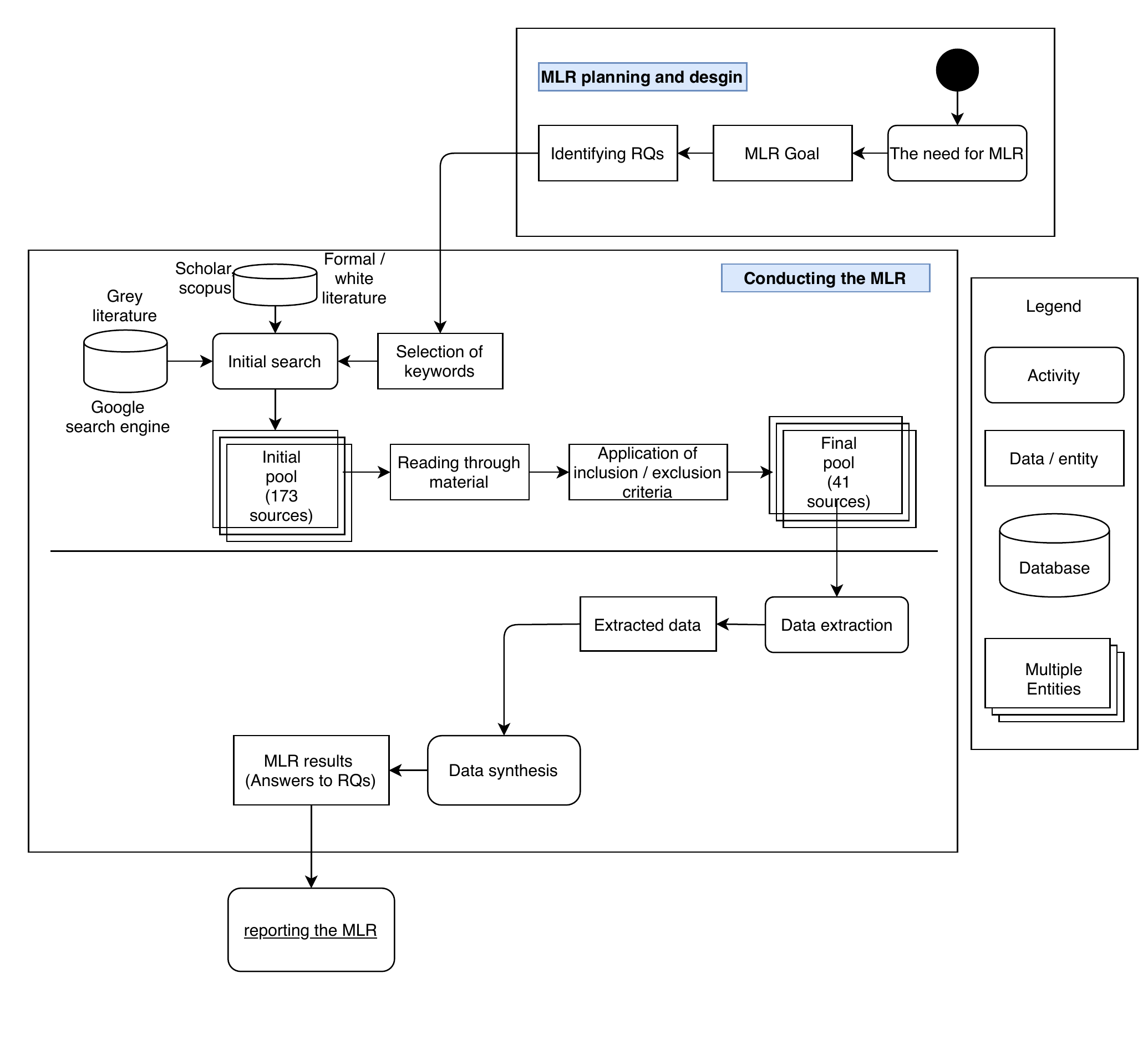}
  \caption{An overview of the described MLR process (as an UML activity diagram).}
  \label{fig:mlrprocess}
\end{figure}

The MLR process adopted was based on five steps (Figure~\ref{fig:mlrprocess}): 
\begin{itemize}[label=$\bullet$]
  \item Selection of keywords and search approach
  \item Initial search and creation of initial pool of sources
  \item Snowballing
  \item Reading through material
  \item Application of inclusion / exclusion criteria
  \item Evaluation of the quality of the grey literature sources
  \item Creation of the final pool of sources
\end{itemize}

The detailed process is depicted in Figure~\ref{fig:mlrprocess}.


\subsection{Search approach}
In this section, we first present the search process adopted for the academic literature, the adaptations we made for the web search, and the snowballing process we adopted.

\subsubsection{Academic Literature Search}
As recommended by Garousi et al~\cite{Garousi18a}, we adopted the traditional Systematic Literature Review process for searching academic literature.

Initially, we selected the relevant bibliographic sources. As including papers from one single publisher may be a bias for an SLR, we considered the papers indexed by several bibliographic sources, namely:
\begin{itemize}
    \item  ACM digital Library \cite{ACMDigital}
    \item  IEEEXplore Digital Library \cite{IEEEXplore}
    \item  Science Direct \cite{ScienceD}
    \item  Scopus \cite{Scopus}
    \item  Google Scholar \cite{GoogleS}
    \item  Citeseer library \cite{CiteseerL}
    \item  Inspec \cite{Inspec}
    \item  Springer link \cite{SpringerL}

\end{itemize}

We adopted  the search strings ''Micro-Frontend*'', ''Micro Frontend*''. Search strings were applied to all the fields (title, abstract, keywords, body, references), so as to include as many academic works of literature as possible. 

The search was conducted in September 2020, and all the raw data are presented in the raw data ~\cite{Peltonen2020}.

\subsubsection{Grey Literature search  }
We adopted the same search strings for retrieving grey literature from Google. 
We applied the Search strings to four Search engines: Google Search, Twitter Search \cite{Reddit}, Reddit Search \cite{Reddit} and Medium Search \cite{Medium}. 


Search results consisted of books, blog posts, forums, websites, videos, white-paper, frameworks, and podcasts. 
Search results from every result page were copied to a Spreadsheet. This search was performed between 16.02.2020 and 17.02.2020. The spreadsheet is available in the replication package~\cite{Peltonen2020}. 




\subsection{Snowballing}
We applied a backwards snowballing to the academic literature, to identify relevant papers from the references of the selected sources. Moreover, we applied backward snowballing for the grey literature following outgoing links of each selected source. 

\subsection{Application of inclusion / exclusion criteria}

Based on SLR guidelines~\cite{Kitchenham2007}, we defined our inclusion criteria, considering academic literature describing the motivation for the adoption of Micro-Frontends, their benefits or issues.

Moreover, we defined our exclusion criteria as: 

\begin{itemize}[label=$\bullet$]
  \item Exclusion criterion 1: Adoption of the term Micro-Frontend for different purposes or different domains (e.g. in mechanics) 
  \item Exclusion criterion 2: Non-English results 
  \item Exclusion criterion 3: Duplicated result
\end{itemize}


\rf[\subsection{Evaluation of the quality and credibility of sources}]

\rf[Differently than peer-reviewed literature, grey literature goes through a formal review process, and therefore its quality is less controlled.  
In order to evaluate the credibility and quality of the selected grey literature sources, and to decide whether to include a grey literature source or not, we extended and applied the quality criteria proposed by Garousi et al~\cite{Garousi18a} (Table~\ref{table:qualityCriteria}), considering the authority of the producer, the methodology applied, objectivity, date, novelty, impact, and  outlet control].

\rf[Two authors assessed each source using the aforementioned criteria, with a binary or 3-point Likert scale, depending in the criteria itself. In case of disagreement, we discussed the evaluation with the third author that helped to provide the final assessment. ]

\rf[We finally calculated the average of the scores and rejected grey literature sources that scored lower than 0.5 on a scale that ranges from 0 to 1.]

\begin{table}[H]
    \centering
    \scriptsize
    \label{table:qualityCriteria}
    \caption{Grey literature quality assessment criteria}
    \begin{tabular}{p{2cm}|p{6cm}|p{6cm}}
\hline
\textbf{Criteria}    & \textbf{Questions} & \textbf{Possible Answers}\\ \hline 
Authority of the producer & Is the publishing organization reputable?	& 1: reputable and well known organization 	\\ \cline{3-3} 
& & 0.5: existing organization but not well known, 0: unknown or low-reputation organization	\\ \cline{2-3}
& Is an individual author associated with a reputable organization?	& 1: true	\\ \cline{3-3}
& & 0: false	\\ \cline{2-3}

& Has the author published other work in the field?	 & 1: Published more than three other work\\  \cline{3-3}
&&0.5: published 1-2 other works, 0: no other works published.	\\  \cline{2-3}

& Does the author have expertise in the area? (e.g., job title principal software engineer)	& 1: author job title is principal software engineer, cloud engineer, front-end developer or similar\\  \cline{3-3} 
&& 0: author job not related to any of the previously mentioned groups. ) \\ \hline 

Methodology & Does the source have a clearly stated aim? & 1: yes\\  \cline{3-3}
&&  0: no	\\ \cline{2-3}

& Is the source supported by authoritative, documented references?	& 1: references pointing to reputable sources \\\cline{3-3}
&&0.5: references to non-highly reputable sources\\  \cline{3-3}
&& 0: no references\\  \cline{2-3}

& Does the work cover a specific question?	& 1: yes\\ \cline{3-3}
&& 0.5: not explicitly\\ \cline{3-3}
&& 0: no\\ \hline 

Objectivity & Does the work seem to be balanced in presentation	& 1: yes\\ \cline{3-3}
&& 0.5: partially\\ \cline{3-3}
&& 0: no\\  \cline{2-3}

& Is the statement in the sources as objective as possible? Or, is the statement a subjective opinion?	 & 1: objective\\ \cline{3-3}
&& 0.5 partially objective\\ \cline{3-3}
&& 0: subjective\\  \cline{2-3}

& Are the conclusions free of bias or is there vested interest? E.g., a tool comparison by authors that are working for particular tool vendor & 1=no interest\\ \cline{3-3}
&& 0.5: partial or small interest\\ \cline{3-3}
&& 0: strong interest\\  \cline{2-3}

& Are the conclusions supported by the data? & 1: yes \\ \cline{3-3}
&& 0.5: partially\\ \cline{3-3}
&& 0: no\\ \hline 

Date & Does the item have a clearly stated date? & 1: yes\\ \cline{3-3}
&& 0: no \\  \hline 

Position w.r.t. related sources & Have key related GL or formal sources been linked to/discussed? & 1: yes\\ \cline{3-3}
&& 0: no \\\hline 

Novelty & Does it enrich or add something unique to the research?	& 1: yes\\ \cline{3-3}
&& 0.5: partially\\ \cline{3-3}
&& 0: no \\\hline 

Outlet type & Outlet Control & 1:  high outlet control/ high credibility: books, magazines, theses, government reports, white papers \\ \cline{3-3}
& & moderate outlet control/ moderate credibility: annual reports, news articles, videos, Q/A sites (such as StackOverflow), wiki articles \\ \cline{3-3}
& &  0: low outlet control/low credibility: blog posts, presentations, emails, tweets \\ \hline 

\end{tabular}
\end{table}


\subsection{Data Extraction and Synthesis}

Based on our RQs, we extracted the information on a structured review spreadsheet. 

To identify motivations,  benefits, and issues we extracted the information from the selected sources via open and selective coding \cite{groundedTheory}. 
The qualitative data analysis has been conducted first by the first author, and then by the last two authors individually.
In a few cases, some motivations, benefits or issues were interpreted differently by some authors. Therefore, we measured pairwise inter-rater reliability across the three sets of decisions and we clarified possible discrepancies and different classifications together, so as to have a 100\% agreement among all the authors.

\subsection{Creation of final pool of sources}
\label{subsec:FinalPoolOfSources}
From the initial pool of 172 sources, 129 sources were excluded. This finalized the pool with 43 sources, from which only 3 (6.97 \%) were  peer-reviewed-conference papers while the other 40 were sourced in the grey literature (e.g., articles, blog posts, videos, books, and podcasts).
\hfill \break


\label{sec:StudyDesign_processDescription}

\section{Study Results and Discussion}
\label{sec:results}

\begin{figure}[H]
  \includegraphics[scale=0.43]{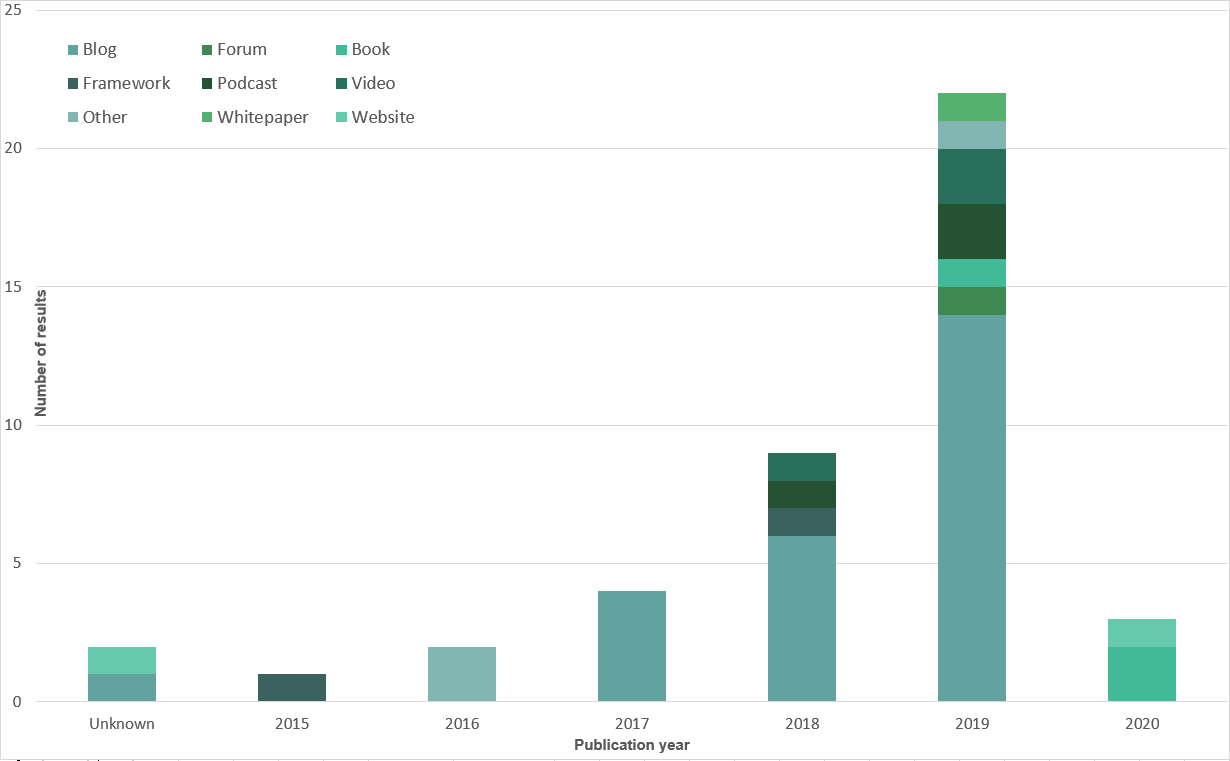}
  \caption{\rf[Distribution of the result  over the years.]}
  \label{fig:ResultHistogram}
\end{figure}

In this section, we present the results of our work, following the research questions presented earlier in Section~\ref{sec:ResearchQuestions}.
The results are based on data extracted from 43 selected sources including 3 peer-reviewed academic paper and 40 Grey Literature sources. 
As we can see from Figure~\ref{fig:ResultHistogram}, the number of publications on Micro-Frontends is constantly growing from 2015. Results from 2020 are lower since the search has been conducted on April 2020. 

\begin{figure}[H]
  \includegraphics[scale=0.43]{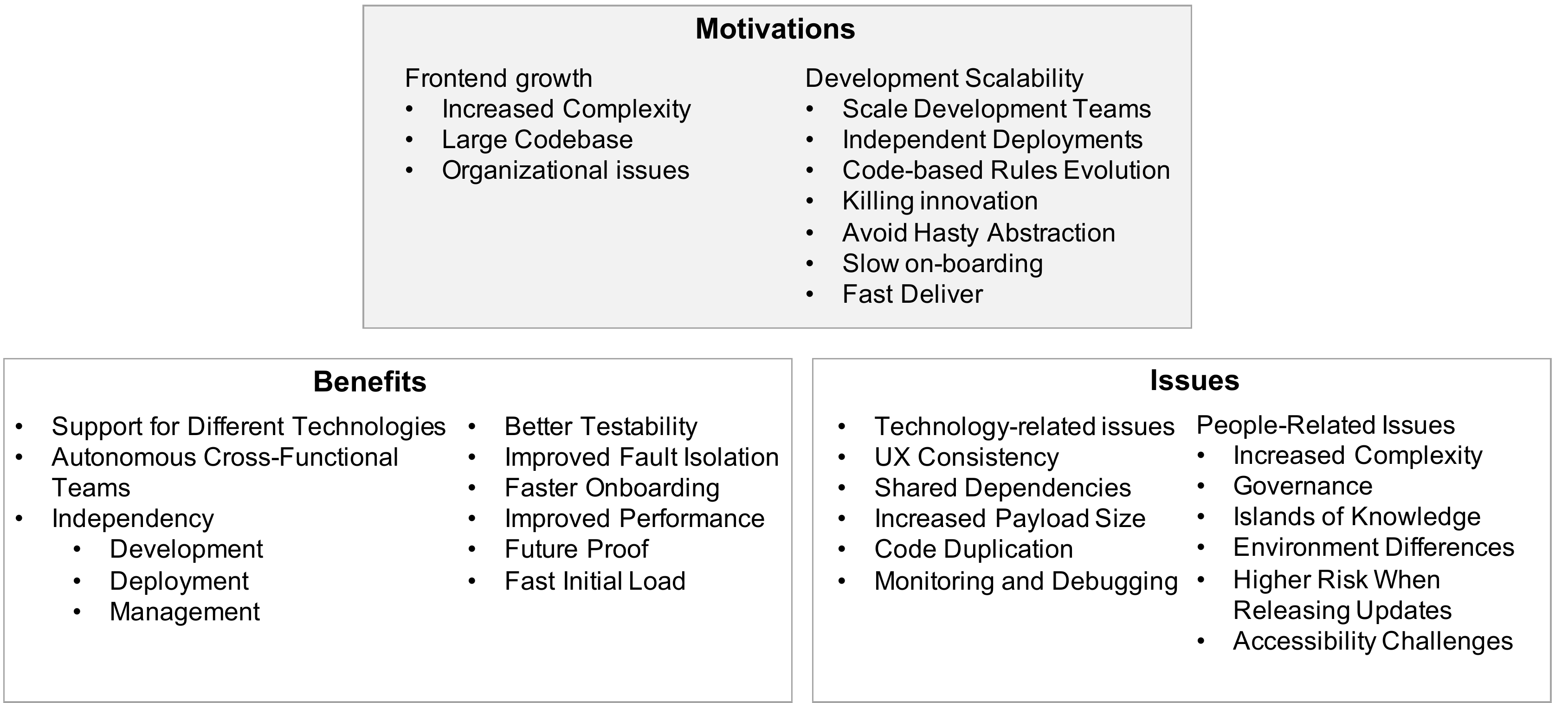}
  \caption{Summary of Motivations, Benefits and Issues}
  \label{fig:Results}
\end{figure}

The results of our RQs are summarized in Figure~\ref{fig:Results}.

\subsection{Why practitioners are adopting Micro-Frontends (RQ1)}
\label{subsec:rq1}

\noindent\textit{Description}. Large companies such as Zalando \ref{Mosaic9}, Ikea \cite{Ikea}, Spotify \cite{Spotify}, and many others are adopting Micro-Frontends. However, the reasons for the adoption are not yet clear to the community. Here we describe and compare the motivations reported by the selected sources. 
\hfill \break

\noindent\textit{Results}. 

The increased complexity of the legacy monolithic frontend and the need to scale the development teams are the main reasons for the adoption of Micro-Frontends. 
Selected sources often mention the problem of delegating responsibilities to independent teams, and the need to ease the support for DevOps. One interesting observation is that several practitioners reported adopting microservices-based architectures and Micro-Frontends because a lot of other companies are adopting them.

In order to quantitatively evaluate the importance of each motivation, in Table~\ref{table:motivationTable}, we report the aforementioned motivations, together with the number of source mentioning them. \textit{Motivation} column summarizes the overall motivation to adopt Micro-Frontends architecture in development processes. Represented motivation is defined after the table. \textit{Sources} column summarizes how many of the result sources mentioned motivation in question and percentage representation out of total sources.



\begin{table}[H]
\centering
\footnotesize
\caption{Motivations for the adoption of Micro-Frontends}
\label{table:motivationTable}
\begin{tabular} {@{}p{8cm}|p{0.3cm}|p{1cm}@{}}
\hline 
\multirow{2}{*}{\textbf{Motivation}} & \multicolumn{2}{c}{\textbf{Sources}} \\ \cline{2-3}
& \# & \% \\ \hline
\multicolumn{3}{c}{\textbf{Frontend growth}} \\ \hline
Increased Complexity  & 16 & 37.21 \\ \hline
Large Codebase & 7 & 16.28 \\ \hline
Organizational Problems & 3 & 6.97 \\ \hline
\multicolumn{3}{c}{\textbf{Development Scalability}} \\ \hline
Need to Scale Development Teams  & 7 & 16.28 \\ \hline
Need of Independent Deployments  & 5 & 11.62 \\ \hline
Code-based Rules Evolution & 4 & 9.30 \\ \hline
Killing innovation & 3 & 6.97 \\ \hline
Avoid Hasty Abstraction  & 2 & 4.65 \\ \hline
Slow on-boarding  & 2 & 4.65 \\ \hline
Fast Delivery  & 1 & 2.32 \\ \hline

\end{tabular}
\end{table}

\begin{itemize}
  \item \textbf {M1: Frontend Growth}

As frontends growth, they become harder and harder to maintain. The growth leads to three motivations for the adoption of Micro-Frontends:
 \begin{itemize}
     \item  \textbf{M1.1: Large Codebase}
        As the backend side of the application has moved 
        to use Microservices, frontend side remains monolithic. 
        Over time front-end grows so big that no team, 
        let alone developer, can understand how 
        the entire application works~\ref{JohnsonMedium18}\ref{Biondic19}.
        Therefore, the monolith frontend becomes hard to scale from the development point of view, and cannot be evolved with current market demands \ref{MoralesMedium18}, \ref{Biondic19}. 
     
    Large applications that are built by using monolith architectures have a lot of dependencies, coordination then becomes harder and more
    time-consuming which leads to the law of diminishing return \ref{Slideshare16}.
    
    The code of each Micro-Frontend will be by definition much smaller than the source code of a single of the monolithic frontend. These smaller codebases tend to be simpler and easier for developers to work with \ref{MartinFowler19}.
    
    While the application grows, there needs to be more developers working with this application. As companies have more developers working on the same team, product managers want to deliver more features, 
this means that the code base is growing fast which imposes a risk \ref{Gohen19}. 

     \item \textbf{M1.2: Increased complexity}
     The front-end will eventually become more and more 
bloated and front-end projects will become more and 
more difficult to maintain and the application becomes
unwieldy~\ref{YangLiuSuIOP19} \ref{GeersMfe19} \ref{PoolaSogeti19} \ref{Spa15} \ref{Slideshare16} \ref{Slideshare16} \ref{xenonstack18} \ref{SoderlundMedium17} \ref{case18} \ref{Krishnamurthy19} \ref{thoughtworksRadar16}.

Every functionality in the application is dependent on each other. 
This means if one function stops working, the whole application 
goes down \ref{ChauhanSoftobiz19}.

    Monolith approach does not allow improvement of software architecture in the long run, software code becomes more abstracted than it should be by increasing code complexity and decreasing its comprehensibility \ref{mezzelira20-2}
    When the project has a medium-large team of developers, all the rules applied to the code base are often decided once at the beginning of the project, and the teams stick with them for months or even years because changing a single decision would require a lot of effort across the entire code base and be a large investment for the organization. \ref{mezzelira20-2}
    As a result, its development complexity rises exponentially with the number of teams modifying it~\ref{Biondic19}\ref{HuangDev19}
    
    Also, the current production application might be done by
last year's tech stack or by code written under a delivery 
pressure, and it's getting to the point where a total rewrite is 
tempting \ref{MartinFowler19}.
     
     \item \textbf{M1.3: Organizational problems}
     Most of the development teams are working in an agile managed project delivery process that advocates cross-functional over a cross-technical team (Angular/React team, Java team, DB team, etc.). Micro-Frontend provides the flexibility to have a cross-functional team over a cross-technical team that focuses on end-to-end delivery \ref{Nitin19}.

    In this common example, naturally, product owners will start to define the stories as front-end and back-end tasks and the cross-functional team will never be a real cross-functional unit. It will be a shallow bubble which looks like an agile team but it will be separated deep inside as features are separated from each other. \ref{ZaferHackernoon19}.
    Chris Coyier says ``Anywhere I’ve worked, these things are a big deal 
and it seems like the industry at large has had endless front-end problems 
with shipping designs that start and stay consistent and cohesive without 
repeating itself with shovelfuls of technical debt.`` \ref{CoyierCssTricks19}.

 \end{itemize}

\item \textbf {M2: Scalability}
\label{item:motivation2}

This work identifies three scalability motivations under comprehensive scalability:

\begin{itemize}
    \item \textbf{M2.1: Need to Scale Development teams}
        Software development processes are complex and often the life cycle of the software can expand from months to even decades. As the life cycle of the application expands the more the application grows over time, the amount of features teams need to support grows also. While multiple teams are contributing to a monolithic application, the more tedious the development and release coordination becomes~\ref{Krishnamurthy19} \ref{Smashing19} \ref{Kumar19} \ref{MartinFowler19} \ref{YangLiuSuIOP19}.
         
    
    Single-page applications, server-side rendering applications or a static HTML page with monolith architecture does not scale well according to business needs because there are not many options to choose from \ref{mezzelira20-2}, this results in the collapse of agile methodologies inside development teams \ref{BrookSmartPate18}.

    \item \textbf{M2.2: Need of Independent deployments}
    
    While updating a monolithic Website or Web application, you need to update it completely. You can’t update just one functionality while keeping the rest of the functionalities old because doing so will cause problems in the website \ref{ChauhanSoftobiz19} \ref{Krishnamurthy19}. This increases
    the chance of breaking the application in production, introducing new bugs and mistakes especially when the code base is not tested extensively~\ref{mezzelira20-2} \ref{PoolaSogeti19}.
    
    With front-end separation to multiple smaller pieces, development teams achieve flexibility in development and operations\ref{Mosaic9}.

    \item \textbf{M2.3: Fast delivery}
    
 The software industry is moving fast forward and companies are dependent on applications and new features on them. These features need to be deployed fast and in a reliable way. With multiple teams working on the same code base this target is hard to achieve \ref{Gohen19}.
        
\end{itemize}

Overall 13  sources (30.95\%) mentioned scalability related motivations for choosing Micro-Frontends architecture.

\item \textbf {M3: Code-base rules evolution}
\label{item:motivation3}

When we have a medium-large team of developers, all the rules applied to the code-base are often decided once, and teams stick with them for months or even years because changing a single decision would require a lot of effort across the entire code-base and be a large investment for the organization. \ref{mezzelira20-2}

\item \textbf {M4: Slow on-boarding}
\label{item:motivation6}

The large code base is confusing and initiation of a new developer is time-consuming because the application has grown too large and has too many edges to explore \ref{Slideshare16}.
Elisabeth Engel describes ``As new developers came to the project and learning it, most of them said that monolith application should be migrated to use more manageable option because understanding the architecture took too long`` \ref{Engel18}.

\item \textbf {M5: Killing innovation}
\label{item:motivation5}

Using a monolithic code base forces developers to introduce new techniques and apply to the entire project for maintaining a code base consistency~\ref{Slideshare16}\ref{PoolaSogeti19}. 
Due to the nature of Micro-Frontends development team can evolve part of the application without affecting the entire system.
In this way, testing a new version of a library or even a completely new UI framework won't provide any harm to the application stability~\ref{Engel18}.


\item \textbf {M6: Avoid Hasty Abstractions}
\label{item:motivation7}

Application has more abstraction layers and top of that is another layer which 
makes the whole architecture complex and more messy \ref{Engel18}.
A lot of code is duplicated and the same thing is done many times over and over again with monolith architectures \ref{Grijsen19}.

Abstractions are always hard to maintain, code duplication, despite it's less elegant, provides greater flexibility and options when we want to refactor.
There are a school of thoughts where a wrong abstraction is way more expensive than code duplicated.
Often developers are abstracting code in components or libraries for using it a couple of times. The problems are not the first iteration where the requirements are clear but the following ones.
In the long run, abstractions may become very hard to maintain and to understand and often not useful at all.
A good technique for implementing the right level of abstraction inside a project is starting with code duplication and when we see duplicated code in more than 3 parts of the applications, try to abstract it.
In this way we keep the flexibility to evolve the code independently, reducing the complexity of abstraction and we are in a position to abstract the code way faster than doing it the other way around\footnote{https://kentcdodds.com/blog/aha-programming \\and https://www.sandimetz.com/blog/2016/1/20/the-wrong-abstraction}.

\end{itemize}

\noindent\textit{Discussion}

As can be seen from results answering RQ1, most of the motivations for the adoption of Micro-Frontends are similar to those for adopting microservices~\cite{TaibiIEEECloud2017}. The increased complexity of the front-end applications often does not allow companies to scale its development processes, assigning different cross-functional features to different teams. Big front-end application also does not allow teams to deliver features fast and decisions made in the beginning of the application development process may end up being unnecessary and done hastily.

\subsection{What benefits are achieved by using Micro-Frontend architecture (RQ 2)}
\noindent\textit{Description}

This section provides results on the benefits that practitioners are receiving by using  Micro-Frontend architectures. \rs[It is interesting to note that all the benefits reported for Micro-Frontends are also common with Microservices~\cite{TaibiIEEECloud2017}\cite{SOLDANI2018}] (Table \ref{table:benefitTable}).

\textit{Results}

Results shown that Micro-Frontends have several benefits such as faster on-boarding for developers, cross-functional teams and imporoved web application performance while they also shares several benefits with microservices such as as they are both technology agnostic, both can be developed, deployed and maintained independently and future proof.

\begin{table}[H]
\centering
\footnotesize
\caption{Micro-Frontends Benefits}
\label{table:benefitTable}
\begin{tabular} {@{}p{9.5cm}|p{0.3cm}|p{1cm}@{}}
\hline 

\multirow{2}{*}{\textbf{Benefit}} & \multicolumn{2}{c}{\textbf{Sources}} \\ \cline{2-3}
& \# & \% \\ \hline
Support for Different Technologies & 22 & 51.16 \\ \hline
Autonomous Cross-Functional Teams &  18 & 41.86 \\ \hline
Independent Development, Deployment and Management  & 15 & 34.88 \\ \hline
Highly Scalable Development  & 5 & 11.63 \\ \hline
Better Testability  & 4 & 9.30 \\ \hline
Improved Fault Isolation, Resiliation & 3 & 6.98 \\ \hline
Faster Onboarding & 3 & 6.98 \\ \hline
Improved Performance  & 2 & 4.65 \\ \hline
Future Proof & 2 & 4.65 \\ \hline
Fast Initial Load & 1 & 2.33\ \\ \hline
\end{tabular}
\end{table}

\begin{itemize}
  \item \textbf {B1: Support for different technologies}
  
With Micro-Frontends, each development team can choose a different technology stack, without the need to coordinate with other development teams
\ref{AgileChamps17}\ref{Krishnamurthy19}\ref{Engel18}\ref{GeersMfe19}\ref{kothariPackt17}\ref{MoralesMedium18}\ref{MotamediRangle19}\ref{PoolaSogeti19} \ref{xenonstack18} \ref{GeersBook20}\ref{Spa15}\ref{Luigi18}\\\ref{Kumar19}\ref{HuangDev19}\ref{Slideshare16}\ref{Mosaic9}\ref{ChauhanSoftobiz19}.
As Micro-Frontend architecture combines multiple
smaller applications into one, the stack and the techniques used will not affect other applications \ref{JohnsonMedium18}\ref{BrookSmartPate18}.

As applications can be implemented in different technologies \ref{Altkom18} — in the world of the rapid evolution of front-end technologies, 
it is impossible to choose an ideal JavaScript framework, which would not be considered as a legacy in the upcoming years.
It is a great benefit that a new framework can be chosen without having to rewrite the existing system.
Technologies can be selected by the development team, based on their needs and their skills \ref{KycPragmatics18}.

\item \textbf {B2: Autonomous cross-functional teams}

Micro-Frontends bring the concept and benefits of Microservices to front-end applications. 
Each Micro-Frontend is self-contained, which allows delivery of fast as multiple teams can work on different parts 
of the application without affecting each other \ref{kothariPackt17}.
Development teams are being able to concentrate on their work without needing permission from the rest of the organization. 
Teams can create innovative architectural decisions inside their applications because the blast radius of those decisions is much smaller \ref{HackerNews19}\ref{xenonstack18}. 

Each team have a distinct area of business that it specializes in \ref{Altkom18}. A team can be cross-functional and develops end-to-end features for large web applications, from the user interface to the back-end and database \ref{MoralesMedium18}\ref{case18}\ref{MotamediRangle19}.  With cross-functional teams, teams have full ownership of everything, from ideation through to production and beyond,  they need to deliver value to customers, which enables them to move quickly and effectively. 
\ref{MartinFowler19}\ref{mezzelira19}\ref{PoolaSogeti19}\ref{ZaferHackernoon19}\ref{Nitin19}\ref{GeersBook20}\ref{Smashing19}\ref{Kumar19}\ref{mezzelira20-2}\ref{MyersToptal}.

\item \textbf {B3: Independent development, deployment and managing and running}

Each Micro-Frontend application is independent. Changing one application will not affect the other parts and it is also more maintainable \ref{Engel18}\ref{PoolaSogeti19}\ref{xteam17}. On a large application, many teams can work parallel and produce features fast without the need to coordinate with other teams \ref{Grijsen19}. Therefore,  Micro-Frontends enable teams to develop
independently, quickly deploy and test individually, helping with continuous integration, continuous
deployment, and continuous delivery \ref{YangLiuSuIOP19}\ref{BrookSmartPate18}\ref{xenonstack18}\ref{KycPragmatics18}\ref{kothariPackt17}\ref{JohnsonMedium18}. 
Since all the front-end modules of the Website or web application are independent of each other, you can develop, test, and deploy them in parallel. 
This reduces the development time and results in faster deployment. \ref{ChauhanSoftobiz19}

By creating small independent applications or modules, resources and teams can proficiently work in separate technologies in their isolated Microservices reducing the risk of conflicts, bugs, and deployment delays \ref{MoralesMedium18}. 


Also, the source code for each Micro-Frontend will by definition be much smaller than the source code of a single monolithic frontend. 
These smaller codebases tend to be simpler and easier for developers to work \ref{MyersToptal}.  Independent deployment reduces the scope of   deployment,
 which in turn reduces the associated risk  \ref{MartinFowler19}.

Known ownership of “verticals” enables better DevOps and faster incident response \ref{Biondic19}

Developers can focus on their work and deliver business value; with less technical synchronization with other teams is needed. 


\item \textbf {B4: Better testability}

Testing becomes simple as
well as for every small change, you don’t have to go and touch the entire application \ref{Krishnamurthy19}\ref{xenonstack18}.

With Micro-Frontends, testing becomes easier because the developer does not have to run the whole test suite every time \ref{Engel18}.

Changing a part in a monolithic application can have multiple side-effects which lead to changing something else in the application.
While the application is specified only to one domain, testing becomes much easier and will not affect the whole application \ref{AgileChamps17}.

\item \textbf {B5: Improved fault isolation, resiliation}

Using Micro-Frontends built with micro-application, one of the biggest benefits of Micro-Frontend over the traditional monolith structure 
is that in case any issue occurs, there’s no need to shut down the entire frontend application to fix it.
If some application fails in run-time the app-shell can detect this and inform the user about the issue \ref{Engel18}\ref{Biondic19}

This is one of the biggest benefits of Micro-Frontend over the traditional monolith structure. In case any issue occurs, there’s no need to shut down the entire frontend to fix it.
Instead, you can fix the module which is having issues while the rest of the app keeps working. \ref{ChauhanSoftobiz19}

\item \textbf {B6: Highly Scalable}

A loosely coupled architecture with established global standards makes it easier to add new features or spin up
teams when needed \ref{Krishnamurthy19}

Divide and distribute the development of end-to-end features to an arbitrary number of teams, 
who can then independently and rapidly develop, deploy, maintain and operate their solutions \ref{Luigi18}. 

No coupling between the frontends means the complexity of the overall system doesn't go up with the amount of them you have and your organization can scale to infinity without increasing coordination \ref{HackerNews19}.

Also, it is easy to spin-off new development teams if needed \ref{Slideshare16}.

Since Micro-Frontend has a modular structure, you can easily upgrade it according to your business needs or market trends.
You don’t need to upgrade the entire front-end. Instead, you can just upgrade the module that’s needed to be up-scaled now and continue updating as your business needs change \ref{ChauhanSoftobiz19}.

\item \textbf {B7: Faster Onboarding}

Enabling teams to on board and deliver quickly \ref{Krishnamurthy19}\ref{Slideshare16}.
Each time a developer joined the development team, they almost immediately understood the system with confidence \ref{Engel18}

\item \textbf {B8: Fast initial load}

Application shell loads micro applications based on the route when the user comes to the web application \ref{Grijsen19}.

\item \textbf {B9: Improved performance}

Since each app is fragmented into its own Micro-Frontend, if a single feature (one micro
frontend) on an enterprise app isn’t loading fast, it won’t affect the performance of the entire application. It also
makes it possible for certain parts of a webpage to load faster, allowing users to interact with the page before all
features are loaded or needed \ref{Krishnamurthy19}

Resulting in faster responses, less code shipped to the browser, and better total load times.\ref{GeersBook20}

\item \textbf {B10: Future proof}

If something new is coming e.g., new framework it can be easily tested and integrated into Micro-Frontends architecture, 
it can easily be abandoned also \ref{Engel18}.
As teams are now free to choose their technology of choice, this makes the application future proof. Teams do not have to invest in only one framework. \ref{Kumar19}

  
\end{itemize}

\textit{Discussion}

As can be seen from results answering RQ2, several of the benefits reflect the motivations for the adoption. As an example, the problem of scaling the development team can be easily tackled with Micro-Frontends. 
However, based on our experience while developing Micro-Frontend based systems,  performances and initial load time depends on how the system is developed, and in particular, on the composition approach adopted (Section 2.2). In general Micro-Frontends 
are not the fastest implementation. As an example, the JAM  stack\footnote{JAM  stack \url{https://jamstack.org/}} is much faster,  because HTML pages contain all the content, and do not need to load dynamically other components.   
However, a Single Page Application might be faster. If developed with the separation of the bundle (e.g. in webpack you can split the JS bundle, so as you download only the beginning of the SPA) and then you load the remaining one (aka code splitting). Server-side/CDN rendering might also be a  fast option, but you need to access to several APIs to compose the page, as pages are not static.
As for the Fast Initial load, the adoption of Micro-Frontends might improve the overall performance, because you only need to load a smaller footprint. The idea of Micro-Frontends is that you load only what you need, not the whole application. As an example, a payment method is usually based on the SDK provided by the payment provider. In a SPA you load the SDK and you download it every time. In Microfrontends, you only download it when you are logged, reducing some KB of memory for non-logged users.

\subsection{Do Micro-Frontends introduce any issues (RQ 3)}

\textit{Description}
While considering issues that practitioners reported while describing Micro-Frontends in the selected works, we highlighted technology-related issues and people-related issues.  Quantitative results are reported in 
Table \ref{table:issuesTable}.

\begin{table}[H]
\footnotesize
\caption{Micro-Frontends Issues}
\label{table:issuesTable}
\centering
\begin{tabular} {@{}p{9.5cm}|p{0.3cm}|p{1cm}@{}}
\hline
\multirow{2}{*}{\textbf{Issues}} & \multicolumn{2}{c}{\textbf{Sources}} \\ \cline{2-3}
& \# & \% \\ \hline

\multicolumn{3}{c}{\textbf{Technology-related issues}} \\ \hline
UX Consistency & 10 & 23.26 \\ \hline
Shared Dependencies  & 7 & 16.28 \\ \hline
Increased Payload Size  & 5 & 11.62 \\ \hline
Code Duplication  & 2 & 4.65 \\ \hline
Monitoring  & 1 & 2.33 \\ \hline
\multicolumn{3}{c}{\textbf{People-Related Issues}} \\ \hline    
Increased Level of Complexity  & 13 & 30.23 \\ \hline
Governance & 1 & 2.33 \\ \hline
Islands of Knowledge & 1 & 2.33 \\ \hline
Environment Differences  & 1 & 2.33 \\ \hline
Higher Risk When Releasing Updates & 1 & 2.33 \\ \hline
Accessibility Challenges  & 1 & 2.33 \\ \hline
\end{tabular}
\end{table}

\textit{Results}

\textbf{Technology-related issues
}

\begin{itemize}
\item \textbf {I1: Increased payload size}
Shipping multiple technology stacks in micro-frameworks has the potential to negatively impact the end-users. 

As a result, if applications are using more than one JS frameworks (for example, 2 applications use Angular and 1 uses React) then the web browser has to fetch a lot of data, with the results of slowing the loading time of the application~\ref{KycPragmatics18}\ref{MotamediRangle19}\ref{Kjartan19}\ref{JohnsonMedium18}\ref{MyersToptal}.

\item \textbf {I2: Code Duplication}

Independently-built JavaScript bundles can cause duplication of common dependencies, increasing the number of bytes applications have to send over the network to end users. 
For example, if every Micro-Frontend includes its own copy of React, then we're forcing our end users to download React n times. 
There is a direct relationship between page performance and user engagement/conversion, and much of the world runs on internet infrastructure much slower than those in 
highly-developed cities are used to, so teams have many reasons to care about download sizes \ref{MartinFowler19}\ref{KBallZendev19}.

\item \textbf {I3: Shared Dependencies}

The dependency redundancy between sub-projects after integration increases the complexity of
management \ref{YangLiuSuIOP19}\ref{Engel18}\ref{Biondic19}.
At the end of the day, Micro-Frontends will have shared dependencies and shared code \ref{xteam17}. This is hard to nail and requires more testing \ref{Gohen19}

When Micro-Frontends are composed in the browser (client-side rendering) there is no singular build process that can optimize and reduplicate shared dependencies. \ref{FeatherBuzzfeed19}\ref{BrookSmartPate18}.

\item \textbf {I4: UX consistency}

The user experience may become a challenge if the autonomous individual teams go with their own direction hence there should be some common medium to ensure UX is not compromised \ref{ZaferHackernoon19}\ref{xteam17}. As new web development frameworks and libraries are being released 
at a brisk pace, the ability to create interoperable.
UX consistency and rich UIs are harder to achieve \ref{Biondic19} and allowing multiple technologies and isolation increases the risk of lack of consistency.\ref{MoralesMedium18}

UI components between frameworks requires building reusable foundational elements which is time-consuming as well. \ref{Krishnamurthy19}

A possible way to increase UX consistency is to use a shared CSS stylesheet, but it means that all applications would depend on one common resource. Is there a better approach? 
The answer is yes and no. There is no perfect solution, but we would recommend using a separate style sheet for each application. Redundancy causes the user to fetch more data, thus impacting application load-time. 
Additionally, components would have to be implemented at least once, which impacts development cost and consistency. The benefit of this approach is independence. This way, we can avoid teams’ synchronization problems during development and deployment. Having a common style guide, designed for example in Zeplin, helps to keep the look and feel consistent (but not identical) across the whole system. 
Alternatively, we could use a common component library included by each application. The disadvantage of this solution is that whenever someone changes the library, they have to ensure that they do not break dependent applications. It would introduce huge inertia. Moreover, a library in most cases can only be used by a single framework. There is no easy way to implement a UI components library, that could be used by the Angular and React app. \ref{KycPragmatics18}


One of the critical problems is standardizing UX principles. A universal solution is to use a style guide, e.g., Bootstrap, Material Design, among others. 
Communication is the key to ensure everything is running smooth, so creating some rules and standards can help minimize conflicts with the diversity of teams working on a product \ref{MoralesMedium18}\ref{Kjartan19}.

\item \textbf {I5: Monitoring}

Tracking and debugging problems across the entire system is complex \ref{KBallZendev19}. 
\item \textbf {I6: Increased level of complexity}

Micro-frontends are not applicable for every application because of their nature and the potential complexity they add at the technical and organizational levels \ref{mezzelira20-2}

This approach can get quite complex if you need to support a large number of significantly different clients implemented with different technologies (e.g. web, native mobile clients, desktop etc.) \ref{Biondic19}\ref{KycPragmatics18}.

The integration of multiple sub-projects application becomes complicated \ref{YangLiuSuIOP19}\ref{Engel18}\ref{Kjartan19}.
As a  distributed architecture, Micro-Frontends will inevitably lead to having more subjects to manage \ref{MartinFowler19}\ref{AgileChamps17}.

This type of architecture requires more initial analysis to understand how everything will work in integration, and how the application can be broken into smaller modules.
Building microservices, and Micro-Frontends introduces significant architectural complexity which will require deeper analysis and quicker interactions. \ref{MoralesMedium18}\ref{MotamediRangle19}\ref{JohnsonMedium18} \ref{PavlovEtAl2020}.
\ref{KBallZendev19}.



\item \textbf {I7: Governance}

The dependency needs to be managed properly. The collaboration becomes a challenge at a time. The multiple
teams working on one product should be aligned and have a common understanding, though when there is a change in
multiple directions in terms of organizational and technology strategy \ref{Krishnamurthy19}.

\item \textbf {I8: Islands of knowledge}

Many cross-functional teams working on the same product, each one working on a different code base and not exposing what they are really doing to other teams. The same implementation will happen over and over again. This is very costly, time-consuming and unnecessary for the companies. \ref{Gohen19}

We recommend using a community of practices, town hall, internal meetups, and a scrum of scrums sessions to overcome this issue.

\item \textbf {I9: Environment differences}

There are risks associated with developing in an environment that is quite different from production. 
If applications development-time container behaves differently than the production one, 
then the team might find that their Micro-Frontend application is broken, or behaves differently when they deploy it to production \ref{MartinFowler19}.

However, there is a way to mitigate this problem following the testing in production mindset where we deploy our new micro-frontends in production with 0 live traffic and the UAT department can test the new module alongside the existing the rest of the application. 
When the tests are satisfying, we can shape traffic to the new micro-frontends either with a small percentage (canary release) or switching the entire traffic (blue-green deployment)

\item \textbf {I10: Higher risk when releasing updates}

Just as teams are able to distribute new changes instantly across many services, They are also able to distribute bugs and errors. 
These errors also surface at application run-time rather than at build time or in continuous integration pipelines. \ref{FeatherBuzzfeed19}

Feature flag and canary releases can help to avoid this problem. If they are totally independent you can only break the single Micro-Frontend. 

\item \textbf {I11: Accessibility challenges}

Some of the implementations of Micro-Frontends, particularly looking at embedding iFrames, can cause huge accessibility challenges\ref{KBallZendev19}. 

Our recommendation, if the application has accessibility requirements it is simply to avoid using iFrames.

\end{itemize}

\textit{\ra[Discussion]}

\ra[Based on the results of our last RQ (RQ3), Micro-Frontends cannot be considered the silver bullet of web frontends. Companies need to take into account the increased complexity introduced by Micro-Frontends, not only from the technical point of view, but also from the management and coordination point of view. Despite Micro-Frontends enable a higher independence between teams, teems need to coordinate and synchronize them to avoid possible inconsistencies of the user interface or of the user experience, and in particular, they need to reduce the need of modifying shared dependencies as much as possible. ]

\ra[Compared to monolithic frontends, debugging is much more complex, especially when testing involve the interactions between different Micro-Frontends. A possible solution might be the development of monitoring tools similar to the ones adopted in serverless functions~\cite{Lenarduzzi2021IEEETest}].
\label{sec:StudyResults}

\section{Discussion}
\rs[In this section, we will discuss the  results obtained  outlining some implications for researchers and practitioners.]
\rs[Although Micro-Frontends are relatively young compared to other web technologies such as static HTML sites, significant contributions have been published by practitioners in the last years (Figure~\ref{fig:ResultHistogram}).]

\rs[Practitioners adopted Micro-Frontends to overcome the issues related to the growth of the monolithic frontend, enabling different teams to develop part of the systems. The adoption of Microfrontend resulted in different benefits, that are also common with microservices, such as the support for different technologies, and the possibility to have autonomous cross-functional teams. ]
\rs[However, this work also enabled us to highlighted some \textit{technology-related issues} and \textit{people-related issues} such as the increased payload size of the frontends, the complexity of the debugging and the need of taking clear architectural and technological decision upfront.]

\rs[Many of the technological issues are mirrored with people-related issues and vice versa, hence when choosing Micro-Frontends architecture development teams need to enhance communication, share knowledge and design well upfront to tackle these issues.]

\rs[Thanks to this work, practitioners will be able to access the main benefits and issues of Micro-Frontends experienced by other practitioners, therefore taking a more thorough decision while deciding if adopt Micro-Frontends or not. ]

\rs[\subsection{Implication for practitioners}
 Software development moves us towards continuous development and delivery. Micro-Frontends support companies in increasing the independence between teams, enabling larger teams, or multiple teams, to work on different part of the frontends simultaneously.  
 However, before deciding to adopt Micro-Frontends, practitioners should take different aspects into account: ]
 
\begin{itemize}
    \item  \rs[Creating a Micro-Frontends application is complex, even more than creating a microservices-based application. Micro-Frontends are not suitable for each and every microservice-based  application. Reasons include the complexity due to the dynamic composition of the web pages, orchestration, testing and debug.]
\item\rs[Working with Micro-Frontends require continuous investments for constantly improving the automation pipeline. Avoiding this investment might impact the speed of delivery for each team working in the project as well as the confidence to deploy in production. ] 
 \item \rs[Micro-frontends increase the payload size, and therefore increase the infrastructure cost. ]
     \item \rs[Micro-Frontend require a thorough architectural design of the system and decision guidance where shared dependencies, and possible duplications should be minimized as much as possible. ] 
 \end{itemize}

\rs[An important observation is that microservices and Micro-Frontends require full application stacks. That means that their infrastructure resources like data stores and networks have to be managed properly. ]

\rs[\subsection{Implications for researchers}
Researchers have not yet deeply investigated Micro-Frontends. ]

\rs[From this work we can highlight the following implications and open issues that researchers might investigate in the future: ]

\begin{itemize}
    \item\rs[ \textit{Comparison between Micro-Frontends and other technologies}. The differences between different technologies and Micro-Frontend (such as single-page-application) has not been thoroughly investigated. On this matter, we can see a lack of comparison from different points of view (e.g., performance, development effort, maintenance).]
    
    \item \rs[\textit{Frontend Duplications}. The practitioners community is now in favour on duplicated code on the frontend. However, no studies investigated and proposed approaches to understand when is appropriate or not to duplicate and abstract the code, and if the duplication of the code in the frontend to different teams is  beneficial or not. From our point of view, we believe that wrong abstractions are more expensive than duplication. 
    However, further works are needed con confirm our hypothesis. ]
    \item \rs[\textit{Patterns and Anti-Patterns}  have not been identified. As an example, there might be different options to connect a Micro-Frontend to multiple microservices, but also to connect multiple Micro-Frontends to each others. ] Researchers might adopt  approaches similar to the ones adopted for microservices~\cite{Neri2020}\cite{Taibi2018closer18}\cite{Taibi2018IEEE}\cite{Taibi2020MSE} and serverless functions~\cite{Taibi2020closer2020}.  
    \item \rs[\textit{When Micro-Frontends are counterproductive?} While some practitioners highlighted that Micro-Frontends are counterproductive for small development teams, other criteria, and factors to decide if adopt Micro-Frontends or not have not yet investigated. ]
    \item \rs[\textit{Combination of Micro-Frontends with other frontend technologies}. It would be interesting to investigate in which condition it might be beneficial to combine Micro-Frontends with other frontend technologies. As an example, large applications might have some part of the frontend developed as static HTML pages, other parts developed as monolithic frontends, and other as Micro-Frontends]
    \item \rs[\textit{Micro-Frontends Migration processes}. While processes to migrate from monolithic systems to microservices have been deeply investigated~\cite{SOLDANI2018}, \cite{TaibiIEEECloud2017}, \cite{Balalaie2018}, process to migrate frontends need to be studies more by researchers. ]
    \item \textit{Micro-Frontends Technical Debt}. As happens when monolithic systems are decomposed into microservices~\cite{Lenarduzzi2020JOURNALA}, decomposing a monolithic frontend into Micro-Frontends might also affect the technical debt. 
\end{itemize}

\label{sec:discussion}

\section{Threats  to Validity}

Our paper might suffer from threats related to the \rs[inaccuracy of the data extraction, a possible  incomplete set of results due to limitation of the search terms, bibliographic sources and grey literature search engine, and  possible subjectivity related to the definition and the application of the  exclusion/inclusion criteria. ]

In the this section, we discuss these threats and the strategies we adopted to mitigate them, based on the standard checklist for validity threats proposed in~\cite{WohlinExperimentation}.

\rs[\subsection{Internal Validity}
The source selection approach adopted in this work  is described in Section~\ref{sec:design}. In order enable the replicability of our work, we carefully identified and reported bibliographic sources adopted to identify the peer-review literature, search engines, adopted for the grey literature,  search strings as well as inclusion and exclusion criteria. ]

\rs[Possible issues in the selection process are related to the selection of search terms that could have lead to a non complete set of results. 
To overcome to mitigate this risk, we applied a broad search string. This was possible because of the novelty of the topic.]

\rs[To overcome the limitation of the search engines,] we queried the academic literature from eight bibliographic sources, while we included the grey literature from Google, Medium Search, Twitter Search and Reddit Search. Additionally, we applied a snowballing process to include all the possible sources.


\rs[The application of  inclusion and exclusion can be affected by researchers’ opinion  and experience. To mitigate this threat, all the sources were evaluated by at least two authors independently. ]



\rs[\subsection{Construct validity}
Construct validities are concerned with issues that to what extent the object of study truly represents theory behind the study~\cite{WohlinExperimentation}. The RQs and the classification schema adopted might might suffer of this threat. To limit this threat, the authors reviewed independently and then discussed collaboratively RQs and the related classification schema.
]

\rs[\subsection{Conclusion validity}
Conclusion validity is related to the reliability of the conclusions drawn from the results~\cite{WohlinExperimentation}.
To ensure the reliability of our treatments, the terminology adopted in the schema has been reviewed by the authors to avoid ambiguities. All primary sources were reviewed by at least two authors to mitigate bias in data extraction and each disagreement was resolved by consensus, involving the third author.]


\rs[\subsection{External Validity}
External validity is related to the generalizability of the  results of our multivocal literature review. ]

\rs[In our study we map the literature on Micro-Frontends, considering both the academic and the grey literature. However, we cannot claim to have screened all the possible literature, since some documents might have not been properly indexed, or possibly copyrighted or, even not freely available. ]

\label{sec:ttv}

\section{Conclusion}
This work presents the results of the first systematic survey on Micro-Frontends, investigating the motivations that led companies to adopt them,  the benefits and issues they experienced.  We conducted a Multivocal Literature Review (MLR)\cite{Garousi18a}, considering 43 sources (3 academic  and 40 grey)  literature, as presented in section \ref{subsec:FinalPoolOfSources}. The main findings of this study confirm that companies and practitioners are seeking alternative architectures for web-frontend development in order to scale development processes and enhance innovation in a rapidly changing business field and allow development teams to be independent and technologically agnostic. 

The most common motivation to adopt Micro-Frontends is the growth of the monolithic frontends (60.46\%) and the consequent increase in code complexity and the need to scale development processes to multiple teams (30.95\%). 

Micro-Frontends architecture provides the same benefits to the frontend side as microservices did to the back-end side of the application. The most mentioned benefits are support for different technologies (50.00\%), Autonomous cross-functional teams (42.86\%) and independent development, deployment, and managing and running (36.71\%).

However, Micro-Frontends are not a silver bullet for designing frontend applications as they increase the overall complexity of the systems (28.57\%) increases payload size (11.90\%), and in general increase development and cloud-related costs.

\rt[This work enabled us to highlight some implications for practitioners and researchers. ]

\rt[As the complexity of the application increases with the Micro-Frontends development teams need to have proper tools to manage the overall complexity in a effective manner. Without effective tools developers have to use a lot of time just to manage the project and this takes time out of the actual feature development. As for now, new tools for creating Micro-Frontends are still merging into the market and teams community still does not have fully ready-to-use solutions that include all the necessary functionality to
support the Micro-Frontends architecture.]

Finally, as this study focused on comprehensive motivation, benefits, and issues with Micro-frontends architecture future examination is required to broaden the scope with the implementation details to find out the advantages and disadvantages of Micro-Frontends composition patterns and overall affection to the development processes.
\rt[Researchers should also support practitioners in understanding the differences between Micro-Frontends and other technologies, and when and why it is beneficial to duplicate the code between teams. Moreover, it is also important to support practitioners in understanding when the adoption of Micro-Frontends can be beneficial  and when it is  counterproductive.]

We are planning to conduct a survey among practitioners to confirm the results obtained in this work and to understand how to properly architect a system based on microservices, Serverless Functions~\cite{Nupponen2020}\cite{Taibi2020IEEE2}, and Micro-Frontends. Future works also include the investigation of benefits and issues of the different Micro-Frontend composition approaches (See Section 2.2) and other composition and architectural approaches to enable multiple teams to work on the same front-end. 
\label{sec:conclusion}

\newpage
\section*{Appendix A: The Selected Sources} 
\label{The Selected Papers}
{\small
  \begin{enumerate}[labelindent=-5pt,label={[S}{\arabic*]}]


\item \label{mezzelira20-2} 	Luca Mezzalira.	"Building Micro-Frontends"  	\\https://www.buildingmicrofrontends.com/,  	2020.  Accessed:2020-05-17
\item \label{Biondic19} 	Denis Biondic. "Building UIs in DevOps / microservices environment part 2— micro-frontends and composite UIs."\\
https://blog.coffeeapplied.com/building-uis-in-devops-microservices-environment-part-2-micro-frontends-and-composite-uis-ab3d4ac394e, 2019.  Accessed:2020-05-17
\item \label{KycPragmatics18} 	Łukasz Kyć	"Independent micro frontends with Single SPA library."  \\	https://blog.pragmatists.com/independent-micro-frontends-with-single-spa-library-a829012dc5be,  	2018.  Accessed:2020-05-17
\item \label{ChauhanSoftobiz19} 	Parijat Chauhan	"Micro Frontend: Extending Microservices to Client-side Development."  	https://www.softobiz.com/micro-frontend-extending-microservices-to-client-side-development/,  	2019.  Accessed:2020-05-17
\item \label{CoyierCssTricks19} 	Chris Coyier	"Micro Frontends." \\
https://css-tricks.com/micro-frontends/,  	2019.  Accessed:2020-05-17
\item \label{JohnsonMedium18} 	Benjamin Johnson	"Exploring micro-frontends."  \\	https://medium.com/@benjamin.d.johnson/exploring-micro-frontends-87a120b3f71c,  	2018.  Accessed:2020-05-17
\item \label{FeatherBuzzfeed19} 	Ian Feather	"Micro Frontends at BuzzFeed." \\  	https://tech.buzzfeed.com/micro-frontends-at-buzzfeed-b8754b31d178,  	2019.  Accessed:2020-05-17
\item \label{GeersMfe19} 	Michael Geers	"Micro Frontends: extending the mircroservice idea to frontend development." \\  	https://micro-frontends.org/,  	2019.  Accessed:2020-05-17
\item \label{HuangDev19} 	Phodal Huang	"Micro-frontend Architecture in Action with six ways."  \\	https://dev.to/phodal/micro-frontend-architecture-in-action-4n60,  	2019.  \\Accessed:2020-05-17
\item \label{KBallZendev19} 	Kevin Ball	"Microfrontends: the good, the bad, and the ugly."  \\	https://zendev.com/2019/06/17/microfrontends-good-bad-ugly.html,  	2019.  \\Accessed:2020-05-17
\item \label{kothariPackt17} 	Amit Kothari	"What is micro frontend?."\\  	https://hub.packtpub.com/what-micro-frontend/,  	2017.  Accessed:2020-05-17
\item \label{MartinFowler19} 	Cam Jackson	"Micro Frontends."\\  	https://martinfowler.com/articles/micro-frontends.html,  	2019.  Accessed:2020-05-17
\item \label{mezzelira19} 	Luca Mezzalira	"Micro-frontends decision framework."  \\	https://lucamezzalira.com/2019/12/22/micro-frontends-decisions-framework/,  	2019.  Accessed:2020-05-17
\item \label{MoralesMedium18} 	Emmanuel Morales	"Micro Front-Ends: Introduction."  \\	https://medium.embengineering.com/micro-front-ends-76171c02ab17,  	2018.  Accessed:2020-05-17
\item \label{MotamediRangle19} 	Ehsan Motamedi	"Five Things to Consider Before Choosing Micro Frontends."  	https://rangle.io/blog/five-things-to-consider-before-choosing-micro-frontends/,  	2019.  Accessed:2020-05-17
\item \label{MyersToptal} 	Bob Myers	"The Strengths and Benefits of Micro Frontends."\\  	https://www.toptal.com/front-end/micro-frontends-strengths-benefits  \\Accessed:2020-05-17
\item \label{PoolaSogeti19} 	Manjuanath Poola	"Micro Frontend Architecture."  	https://labs.sogeti.com/micro-frontend-architecture/,  	2019.  Accessed:2020-05-17
\item \label{BrookSmartPate18} 	Paul Brook	"Microservice Front-end - A Modern Approach to the Division of the front."  	https://www.smartspate.com/microservice-front-end/,  	2018.  Accessed:2020-05-17
\item \label{xenonstack18} 	-	"Micro Frontend Architecture and Best Practices."\\  	https://www.xenonstack.com/insights/what-is-micro-frontend/,  	2018.  Accessed:2020-05-17
\item \label{ZaferHackernoon19} 	Öner Zafer	"Understanding Micro Frontends."\\  	https://hackernoon.com/understanding-micro-frontends-b1c11585a297,  	2019.  Accessed:2020-05-17
\item \label{Nitin19} 	Nitin Jain	"Micro Frontend Curry."\\  	https://levelup.gitconnected.com/micro-frontend-curry-506b98a4cfc0,  	2019. \\ Accessed:2020-05-17
\item \label{Kjartan19} 	Kjartan Rekdal Müller	"Easy Micro-Frontends. "\\ 	https://itnext.io/prototyping-micro-frontends-d03397c5f770,  	2019.  Accessed:2020-05-17
\item \label{xteam17} 	Juan E. Jimenez	"A take on Micro-frontends. "\\ 	https://x-team.com/blog/micro-frontend/,  	2017.  Accessed:2020-05-17
\item \label{AgileChamps17} 	-	"Microservices to Micro-Frontends."\\  	http://www.agilechamps.com/microservices-to-micro-frontends/,  	2017.  Accessed:2020-05-17
\item \label{Altkom18} 	Roberto Witkowski	"UI in Microservices World – Micro Frontends pattern and Web Components."\\  	https://altkomsoftware.pl/en/blog/ui-in-microservices-world/,  	2018.  Accessed:2020-05-17
\item \label{GeersBook20} 	Michael Geers	"Micro Frontends in Action. "\\ 	https://www.manning.com/books/micro-frontends-in-action,  	2020.  \\Accessed:2020-05-17
\item \label{Kumar19} 	Ajay Kumar	"Micro Frontends Architecture."\\	https://www.amazon.com/Micro-Frontends-Architecture-Introduction-Techniques/dp/1097927989,  	2019.  Accessed:2020-05-17
\item \label{SoderlundMedium17} 	Tom Söderlund	"Micro frontends—a microservice approach to front-end web development."\\  	https://medium.com/@tomsoderlund/micro-frontends-a-microservice-approach-to-front-end-web-development-f325ebdadc16,  	2017.  Accessed:2020-05-17
\item \label{HackerNews19} 	"Micro Frontends."  \\	https://news.ycombinator.com/item?id=20148308,  	2019.  Accessed:2020-05-17
\item \label{Luigi18} 	"The Enterprise-Ready Micro Frontend Framework."\\  	https://luigi-project.io/about,  	2018.  Accessed:2020-05-17
\item \label{Spa15} "Single-spa: A javascript router for front-end microservices."\\  	https://single-spa.js.org/,  	2015.  Accessed:2020-05-17
\item \label{Slideshare16} 	Luiz Mineiro "The frontend taboo."\\  	https://www.slideshare.net/lmineiro/the-frontend-taboo-a-story-of-full-stack-microservices,  	2016.  Accessed:2020-05-17
\item \label{Smashing19} 	Luca Mezzalira	"Smashing Podcast Episode 6: What Are Micro-Frontends?."\\   Smassing Magazine 	https://www.smashingmagazine.com/2019/12/smashing-podcast-episode-6/,  	2019.  Accessed:2020-05-17
\item \label{case18} 	Gustaf Nilsson Kotte	"Micro Frontends with Gustaf Nilsson Kotte."  \\	https://www.case-podcast.org/22-micro-frontends-with-gustaf-nilsson-kotte,  	2018.  Accessed:2020-05-17
\item \label{Thoughtworks19} 	Thoughtswork podacast	"What's cool about micro frontends." \\	https://thoughtworks.libsyn.com/whats-so-cool-about-micro-frontends,  	2019.  Accessed:2020-05-17
\item \label{thoughtworksRadar16} 	Thoughtworks	"Technology radar: Micro frontends."  \\	https://www.thoughtworks.com/radar/techniques/micro-frontends,  	2016.  \\Accessed:2020-05-17
\item \label{Grijsen19} 	Erik Grijzen	"Micro Frontend Architecture Building an Extensible UI Platform."\\  	https://www.youtube.com/watch?v=9Xo-rGUq-6E,  	2019.  Accessed:2020-05-17
\item \label{Gohen19} 	Liron Cohen	"Micro-frontends: Is it a Silver Bullet?" | React Next 2019. \\ 	https://www.youtube.com/watch?v=asqgKaUMXq0,  	2019.  Accessed:2020-05-17
\item \label{Engel18} 	Elisabeth Engel	"Break Up With Your Frontend Monolith" JSCamp Barcelona 2018.  https://tinyurl.com/y5ksc2fm,  	2018.  Accessed:2020-05-17
\item \label{Mosaic9} 	Zalando	"Project Mosaic | Microservices for the Frontend."  	\\https://www.mosaic9.org/,  	2016.  Accessed:2020-05-17
\item \label{Krishnamurthy19} 	Santhosh Krishnamurthy	"What are Micro Frontends."  IEEE India Info. Vol. 14, No. 4, Oct-Dec 2019	
\item \label{YangLiuSuIOP19} 	Caifang Yang, Chuanchang Liu, Zhiuaun Su	"Research and Application of Micro Frontends." IOP Conference on Materials Science and Engineering. 2019. DOI: 10.1088/1757-899x/490/6/062082\\
\item \label{PavlovEtAl2020} Andrey Pavlenko, Nursultan Askarbekuly, Swati Megha, Manuel Mazzara
"Micro-frontends: application of microservices to web front-ends. "\\ Journal of Internet Services and Information Security (JISIS), volume: 10, number: 2 (May 2020), pp. 49-66 DOI: 10.22667/JISIS.2020.05.31.049
\end{enumerate}
}

\newpage
\bibliographystyle{model1-num-names}
\bibliography{bibliography}

%
%
%
%






\end{document}